\documentclass[manuscript]{aastex}

\slugcomment{Not to appear in Nonlearned J., 45.}

\shorttitle{Li abundance of PMS stars in NGC 2264}
\shortauthors{Lim et al.}

\begin{document}


\title{A CONSTRAINT ON THE FORMATION TIMESCALE OF THE YOUNG OPEN CLUSTER NGC 2264:  \\
    LITHIUM ABUNDANCE OF PRE-MAIN SEQUENCE STARS}


\author{Beomdu Lim \altaffilmark{1,6}, Hwankyung Sung\altaffilmark{2}, Jinyoung S. Kim\altaffilmark{3}, Michael S. Bessell\altaffilmark{4},
Narae Hwang\altaffilmark{1}, and Byeong-Gon Park\altaffilmark{1},\altaffilmark{5}}
\email{bdlim1210@kasi.re.kr}

\altaffiltext{1}{Korea Astronomy and Space Science Institute, 776 Daedeokdae-ro, Yuseong-gu, Daejeon 34055, Korea}
\altaffiltext{2}{Department of Astronomy and Space Science, Sejong University, 209 Neungdong-ro, Gwangjin-gu, Seoul 05006, Korea}
\altaffiltext{3}{Steward Observatory, University of Arizona, 933 N. Cherry Ave. Tucson, AZ 85721-0065, USA}
\altaffiltext{4}{Research School of Astronomy and Astrophysics, Australian National University, Canberra, ACT 2611, Australia}
\altaffiltext{5}{Astronomy and Space Science Major, University of Science and Technology, 217 Gajeong-ro, Yuseong-gu, Daejeon 34113, Korea}


\altaffiltext{6}{Corresponding author}


\begin{abstract}
The timescale of cluster formation is an essential parameter in order to understand 
the formation process of star clusters. Pre-main sequence (PMS) stars in nearby young 
open clusters reveal a large spread in brightness. If the spread were considered as a result of 
a real spread in age, the corresponding cluster formation timescale would 
be about 5 -- 20 Myr. Hence it could be interpreted that star formation in an open cluster is 
prolonged for up to a few tens of Myr. However, difficulties in reddening correction, observational 
errors, and systematic uncertainties introduced by imperfect evolutionary models for PMS 
stars, can result in an artificial age spread. Alternatively, we can utilize Li abundance as a relative 
age indicator of PMS star to determine the cluster formation timescale. The 
optical spectra of 134 PMS stars in \objectname{NGC 2264} have been 
obtained with MMT/Hectochelle. The equivalent widths have been measured for 86 
PMS stars with a detectable Li line (3500 $< T_{\mathrm{eff}} [\mathrm{K}] \leq 6500$). 
Li abundance under the condition of local thermodynamic equilibrium (LTE) was derived using the 
conventional curve of growth method. After correction for non-LTE effects, we find that 
the initial Li abundance of \objectname{NGC 2264} is $A\mathrm{(Li)} = 3.2 \pm 0.2$. 
From the distribution of the Li abundances, the underlying age spread of the visible PMS stars is estimated 
to be about 3 -- 4  Myr and this, together with the presence of embedded populations in 
\objectname{NGC 2264}, suggests that the cluster formed on a timescale shorter than 5 Myr.

\end{abstract}

\keywords{stars: formation --- stars: abundance --- stars: activity --- stars: pre-main 
sequence  --- open clusters and associations: individual(NGC 2264)}

\section{INTRODUCTION}
Theories of star formation have predicted that star formation takes 
place on different timescales depending on the physical states of the molecular cloud. 
As reviewed by \citet{SAL87}, star formation in a magnetically subcritical cloud 
can be initiated at a slow rate through ambipolar diffusion. In this scenario, the 
molecular cloud is supposed to be in an equilibrium state where the magnetic field pressure 
supports the cloud against gravitational collapse, and therefore the magnetic 
field plays a crucial role in regulating star formation. Ambipolar diffusion occurs on 
a timescale of 5 -- 10 Myr \citep{TM04} when the density increases up to $n \sim 
10^5 \ \mathrm{cm}^{-3}$ \citep{MPF85}. According to the rapid star formation scenario, 
a giant molecular cloud can be formed on a timescale of a few Myr, and subsequently 
cores are formed by the dissipation of supersonic turbulence inside the cloud \citep{BHV99,E00}. 
Star formation is eventually initiated in the collapsing cores. 

It is known that most of stars (about 80 -- 90\% ) in a star forming region constitute 
a cluster \citep{LL03,PCA03}. After the onset of star formation, the duration of star 
formation activity within a protocluster is still poorly constrained. There are two 
questions: (1) Are low density associations and gravitationally bound clusters formed 
on a different timescale? (2) Can observational properties, such as the extended main 
sequence (MS) turn-off of several star clusters in the Large Magellanic Cloud \citep[etc]{MB07,MBPA09}
and multiple stellar populations in Galactic globular clusters \citep[etc]{PMA12,VDD14} 
be explained by extended star formation processes? The duration of star formation activity 
is an essential quantity for drawing a general picture of cluster formation.

If the age distribution of cluster members can be reliably obtained, the age spread among 
them gives the formation timescale of the cluster. The star formation 
history of nearby young open clusters has been investigated by a large number of 
previous investigators from the color-magnitude diagrams (CMDs). \citet{H62} found a large age 
difference between the MS turn-off and the MS turn-on from the pre-main sequence (PMS) stage 
in the Pleiades. He presumed that low-mass star formation occurred prior to the formation of high-mass 
stars, and that star formation within a cluster could be prolonged for up to $\sim 10^8$ years. 

\citet{PaSt99,PaSt00,PaSt02} suggested a revised version of the sequential 
star formation speculated by \citet{H62} using their own PMS evolutionary models. According 
to their argument, star formation started at an early epoch (about 10 Myr ago), 
and then the star formation rate accelerated in the present epoch (1 -- 2 Myr ago). 
\citet{PRFP05} strengthened their idea by invoking the significant Li depletion of four PMS 
members of the Orion Nebula Cluster. It meant that the formation timescale of a cluster 
is about 10 Myr. Recently, the star formation history of the starburst 
cluster NGC 2070 in the Large Magellanic Cloud was studied by \citet{CSv15}, and the result 
corresponds well to the extended star formation scenario suggested by \citet{PaSt99}. 
\citet{SBL97,SBC04} raised doubts on the sequential star formation scenario from 
the age distribution of PMS stars in NGC 2264, although they found a spread in age among 
PMS stars with H$\alpha$ emission. \citet{H99,H01,H03} also argued that 
various factors, such as binarity, variability, extinction, accretion activities, the inclusion 
of field stars, and uncertainties in the evolutionary models for intermediate-mass PMS stars, 
are responsible for the apparent age spread. Although many follow-up studies have 
examined such factors, the issue of the age spread among PMS stars in a cluster is 
still under debate.

The extended star formation scenario introduced above is based on the analysis of the 
CMDs with various evolutionary models for PMS stars. As mentioned by \citet{H99,H01,H03}, large 
observational uncertainties may be involved in the luminosities of PMS stars, and it is 
difficult to control the sources of these uncertainties from a single observation. The modulation of 
spots, the obscuration by disk material, an edge-on disk, and accretion activities may be typical 
sources of the observational uncertainties \citep{SSB09,SCR16}. In addition, as most young stars 
are deeply embedded in the natal molecular cloud, the difficulty of reddening correction of low-mass 
PMS stars and differential reddening across the cluster, are other sources of scatter in the 
CMDs. Hence, independent age indicators are required to disentangle this issue.

 Lithium (Li) is a fragile light element that is destroyed at a temperature of about 
$\sim 2 - 3 \times 10^6$ K. The temperature  at the base of convective envelope 
inside a low-mass PMS star easily reaches the destruction temperature as the star 
contracts \citep{Br65,PT02}. Li-depleted material at the base of the convective 
envelope is transported to the surface, and therefore in a slow-rotating low-mass 
PMS star, the Li abundance on the surface decreases steadily over time. The 
surface Li abundance has long been regarded as a good stellar age indicator 
and as a diagnostic tool to probe the stellar interior structure \citep{H65,P10}.

A number of previous studies focused on the Li abundance of low-mass stars in nearby open 
clusters to understand the process of Li depletion since \citet{H65} recognized a 
correlation between Li abundance and stellar age. \citet{WHC65} supported Herbig's suggestion 
by invoking the higher Li abundance of a few Hyades members compared to that of the Sun. They also 
suggested that the low Li abundance of old cool stars may be attributed to the Li depletion process 
during the PMS contraction phase. \citet{Z72} observationally tested the theoretical 
Li depletion model of \citet{Br65} for PMS stars, and argued that an extra depletion on a timescale 
of about 1 Gyr during MS evolution (in the MS band) was required to explain the Li 
abundance of the solar-type Hyades members. \citet{D81} reported the Li deficiency of a few F-type stars in the Hyades. 
This feature, the so-called `Boesgaard funnel' or `chasm', turned out to be common among 
F-type stars in intermediate-age clusters \citep{BT86,SOJ90,SJB93,SPFJ93}, and it seems to be 
associated with rapid rotation \citep{Bd87}. 

While cooler members of the Hyades showed only a small scatter in Li abundance, a large 
intrinsic scatter was found for Pleiades members. \citet{DJ83} presumed that the observed 
scatter may be due to an age spread among members. \citet{SJB93} found most
Li-rich stars to be fast rotators, and suggested that the spread in the Li abundance may be 
related to a difference in stellar rotation. \citet{SKSJF99} investigated the rotational 
period and Li abundance of PMS stars in \objectname{NGC 2264} as a younger counterpart of 
the Pleiades, but no meaningful correlation was found. Recently, \citet{BLV16} reported 
only a weak correlation between the Li equivalent width [$W_{\lambda6708}$(Li)] and rotational 
velocity in the same cluster. These facts may imply that the effects of stellar rotation on the 
Li abundance may be less significant for 3 -- 5 Myr young stars.  

The strength of the Li {\scriptsize \textsc{I}} line in cool stars is very sensitive to temperature, and therefore spots 
on the surface and chromospheric activity could be other sources causing a variation in Li abundance. Many 
previous studies tried to confirm the correlation between Li abundance and stellar activity, but little 
variation in the strength of Li {\scriptsize \textsc{I}} $\lambda$6708 doublet was found \citep[and 
references therein]{PCRG93,M98,NGSG01}. A recent study of the young active star 
\objectname{LQ Hydrae} showed a small variation of Li abundance (about 0.02 dex), which was 
synchronized with the rotational modulation \citep{SSW15}. 

\begin{figure}
\epsscale{.80}
\plotone{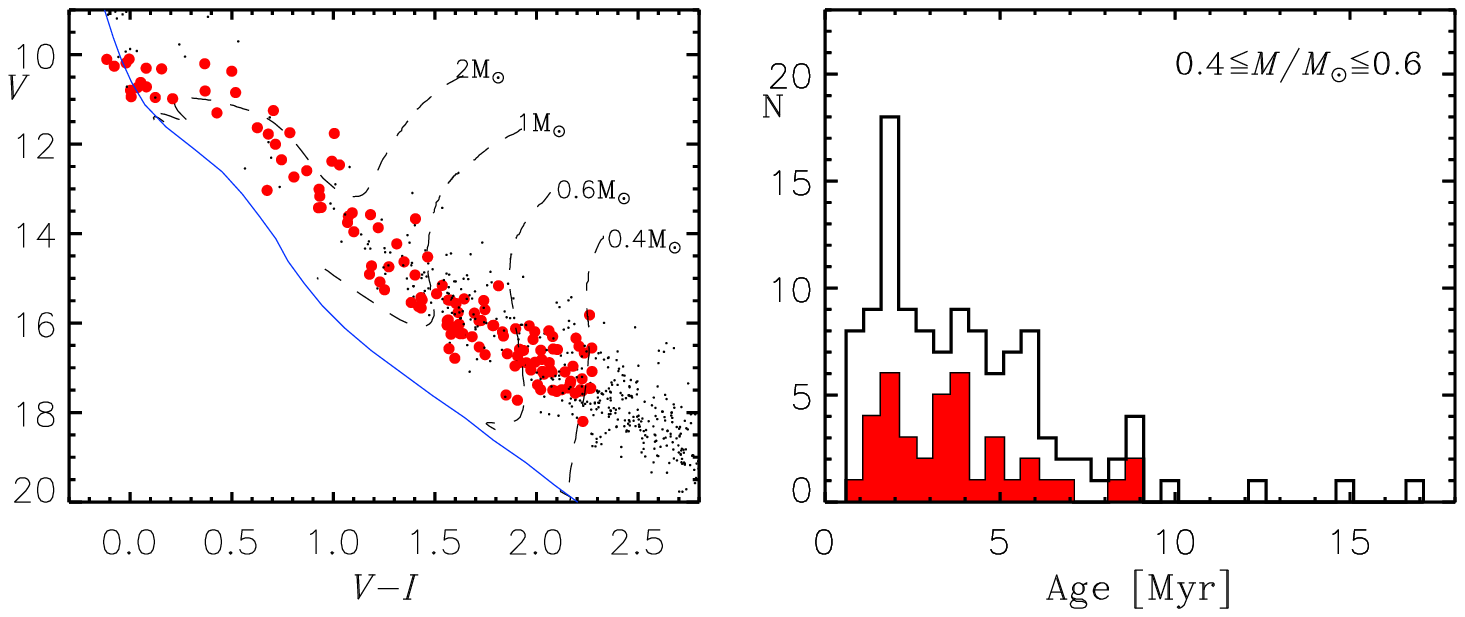}
\caption{Color-magnitude diagram of all known pre-main sequence (PMS) members of NGC 2264 
\citep{SBL97,PSBK00,SBC04,SBCKI08,SSB09} (left) and their age distribution (right). The stars 
observed for this study are marked by bold red dots. Solid and dashed lines represent the zero-age main 
sequence relation and PMS evolutionary tracks for stars of different mass \citep{SDF00}, 
respectively. The open histogram is the age distribution of all the PMS members in the mass 
range of 0.4 $M_{\sun}$ to 0.6 $M_{\sun}$, while the shaded histogram is that of all observed 
stars}\label{fig1}
\end{figure}

\objectname{NGC 2264} is one of nearby young open clusters ($d = 740 - 760$ pc -- 
\citealt{SBL97,KIO14}). More than one thousand stars have been identified as members of the cluster 
\citep{SBL97,PSBK00,SBC04,SBCKI08,SSB09}. Photometric studies have indicated a large 
apparent age spread among the PMS members as shown in Figure~\ref{fig1}. The spread in age 
of PMS stars in the cluster was also reported in other studies \citep{FMS99,PaSt00,DS05}. 
Therefore, this cluster is an ideal test bed to either prove or disprove the apparent age 
spread. We will utilize the Li abundance as a relative age indicator among PMS stars within the cluster. 
Since it can be assumed that all members were formed from the same molecular cloud, 
this analysis is free from significant star-to-star differences in the initial 
chemical composition \citep{FK14}. In addition, \objectname{NGC 2264} is so young that 
the Li abundance of low-mass PMS stars may be less affected by rotationally-induced mixing 
arising from angular momentum loss. 

The observations and data reduction are described in 
Section 2. In Section 3, we discuss the equivalent width measurements of Li {\scriptsize \textsc{I}} 
$\lambda$6708, and derive the Li abundances. The age 
spread and formation timescale of the cluster are derived in Section 4. A few Li deficient 
stars are discussed in Section 5. Finally, we present a summary and conclusions of this 
study in Section 6.

\section{OBSERVATIONS AND DATA REDUCTION}
Membership selection in open clusters is a crucial issue for obtaining reliable scientific 
results. \citet{SB10} carefully investigated the nature of PMS stars in \objectname{NGC 2264} 
using their member list previously obtained from a variety of membership criteria (e.g., X-ray, 
ultra-violet excess, H$\alpha$, and mid-infrared excess emissions). We used their 
member list to select spectroscopic targets. Figure~\ref{fig1} shows the ($V$, $V-I$) 
CMD of the member stars and their age distribution from the CMD. The current evolutionary models 
for PMS stars predict that stars in the mass range of 0.4 $M_{\sun}$ to 0.6 $M_{\sun}$ 
completely deplete their initial Li within 15 Myr \citep{BHAC15}. Therefore, the Li abundance of 
these subsolar mass stars could provide a reliable age indicator, at least on a relative scale. 
In order to determine the initial Li abundance of \objectname{NGC 2264}, higher 
mass PMS stars were also included in our target list.

Queue scheduled observations were carried out on 2015 April 1 and November 24 with 
the multi-object high resolution echelle spectrograph Hectochelle \citep{SFC11} attached to the 6.5 m 
telescope of the MMT observatory. The resolving power of the spectrograph ($R \sim 34,000$) 
is high enough to detect the  Li {\scriptsize \textsc{I}} $\lambda 6708$ resonance doublet
with little blending from adjacent metallic lines. The multi-object capability 
allowed us to simultaneously obtain 240 target and sky spectra in a single observation. The OB 26 
filter transmits the wavelength range 6530 -- 6715 \AA \ \citep{SFC11}, and therefore the 
useful spectral features H$\alpha$ $\lambda 6563$ and He {\scriptsize \textsc{I}} $\lambda 
6678$ could also be observed along with the Li {\scriptsize \textsc{I}} $\lambda 6708$ line. The 
spectra of a total of 134 PMS stars were taken in two sets of exposure 
times -- 8 minutes $\times$ 3 for bright stars ($V < 13.6$ mag), and 30 minutes $\times$ 3 for 
fainter stars. Offset sky spectra were also obtained to correct for the contributions of locally 
variable nebula emission lines to the spectra of the faint stars. Calibration frames, such as dome 
flat and comparison spectra were also acquired, just before and after the target exposure.

The multiple extension FITS (Flexible Image Transport System) images were merged after overscan correction
into a single FITS image using \textsc{IRAF/MSCRED} packages. 
One-dimensional (1D) spectra of PMS stars were extracted using the {\it dofibers} task 
in the \textsc{IRAF/SPECRED} package. A total of 240 apertures were 
traced in the dome flat spectra using spline function fitting, and the 1D spectra of all the fibers 
were extracted along each aperture. The pixel-to-pixel variation was also corrected for
using the residuals of the dome flat spectra divided by a high-order spline function. We 
determined the solutions for the wavelength calibration using ThAr lamp spectra, and 
applied them to the target and flat spectra. In order to eliminate the blaze function of 
the echelle spectra, the target spectra were divided by the wavelength-calibrated dome 
flat spectra.

\begin{figure}
\epsscale{.80}
\plotone{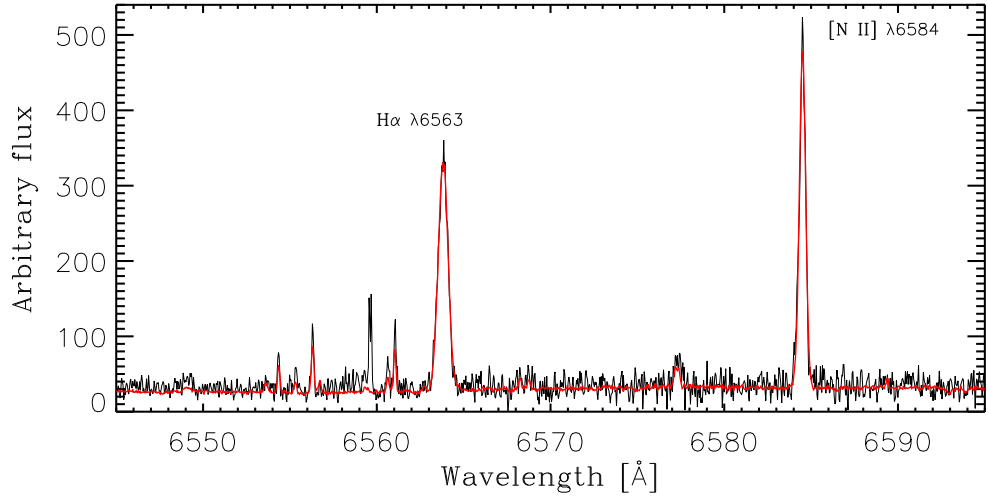}
\caption{Comparison of an offset sky spectrum (black) with a synthetic one (red). Prominent nebula emission 
lines are located between 6540 -- 6590 \AA \ in our echelle spectrum for the OB 26 filter. The emission lines 
H$\alpha$ and [N {\scriptsize \textsc{II}}] $\lambda$6584 are identified at the top of the figure. }\label{fig2}
\end{figure}

An offset sky observation was not made for bright stars ($V < 13.6$ mag). Instead, 
we assigned a few tens of fibers to blank sky to obtain the simultaneous sky spectra. A master 
sky spectrum with high signal-to-noise ratio was obtained by median combining those sky 
spectra which were not affected by scattered light from nearby bright stars. The nebular contribution 
to the target spectra were subtracted using the master sky spectrum. As the contribution of locally 
variable emission lines, such as H$\alpha$ $\lambda$6563, [N {\scriptsize \textsc{II}}] 
$\lambda$6548, and $\lambda$6584, becomes significant for faint PMS stars, we created an 
adaptable synthetic sky spectrum following the procedure described by \citet{KKBSR15} 
for the correction of spatially varying nebular lines. Sky spectra not affected by the illumination 
of nearby bright stars were selected from the offset sky observation, and were then combined 
into a master sky spectrum. The nebula emission lines were removed from the master spectrum. 
We extracted the emission lines from each offset sky spectrum corresponding to a given target, 
and then added them to the emission-subtracted master sky spectrum. Figure~\ref{fig2} shows a 
comparison of an offset sky spectrum with a synthetic one. The [N {\scriptsize \textsc{II}}] 
$\lambda$6548 line was not considered in this work because its strength was comparable 
to the noise level of target spectra. The observed spectra of faint PMS stars were corrected by subtracting
the corresponding synthetic sky spectrum.

We found non-negligible residuals of the nebula emission lines in some target spectra. 
Since low-mass active PMS stars, in general, show a broad, strong H$\alpha$ emission line, 
it is sometimes difficult to directly distinguish the H$\alpha$ emission line of the nebula from that of the target 
spectra. However, the [N {\scriptsize \textsc{II}}] $\lambda$6584 line is strong and well isolated, so
assuming the ratio of the [N {\scriptsize \textsc{II}}] $\lambda$6584 line to the H$\alpha$ 
emission line is constant within a small local area, we scaled the strength of the two emission lines 
in the synthetic sky spectrum to that in a given target spectrum by comparing the [N {\scriptsize \textsc{II}}] 
$\lambda$6584 line strength in each spectrum. The fine-tuned sky spectrum was then subtracted 
from the target spectrum, and the sky-subtracted target spectra, for the same target, were 
combined into a single spectrum. Finally, the individual target spectra were normalized 
using the solution found from a cubic spline interpolation to the continuum level.

\section{SPECTRAL LINE ANALYSIS}

\subsection{Estimation of Effective Temperature}

The effective temperature is one of the key stellar parameters in the analysis of the spectral lines. 
Color-temperature relations have been used to estimate the effective temperature of the target 
stars in a convenient way. The application of this method to the stars 
in \objectname{NGC 2264} has an advantage because the amount of interstellar reddening 
toward the optically visible stars in the cluster ($\langle E(B-V)\rangle = 0.07 \pm 0.03$ mag -- 
\citealt{SBL97}) is low compared to that toward other clusters in different spiral arms 
($\langle E(B-V)\rangle = 4.2$ mag for \objectname{Westerlund 1} in the Scutum-Centaurus arm 
-- \citealt{LCS13}; 0.5 mag for \objectname{NGC 6231} in the Sagittarius-Carina arm -- \citealt{SSB13}; 
0.6 mag for \objectname{NGC 1893} in the Perseus arm -- \citealt{LSKBP14}; 0.8 mag for 
\objectname{IC 1805} in the Perseus arm -- Sung et al. in preparation; 0.9 mag for \objectname{NGC 1624} in 
Outer arm -- \citealt{LSB15}). In addition, differential reddening is not severe among the massive 
stars. 

Spectral features caused by accretion activities can also influence the observed colors. The 
standard accretion model demonstrates that accretion flows from a circumstellar disk 
fall onto the stellar surface along the magnetosphere \citep{US85,BBB88,K91}. Hot continuum 
excess emission and a large variety of recombination lines driven by the accretion process 
are observed in many young low-mass PMS stars \citep{CG98,H99}. Colors, such as $U-B$ 
and $B-V$, can be affected, particularly by the hot continuum excess emission. On the other 
hand, the $V-I$ color is less sensitive to the effects of accretion \citep{SBL97}, and the color also has 
higher temperature resolution than $B-V$ for cool stars. For this reason, we applied the 
$V-I$ versus $T_{\mathrm{eff}}$ relation of \citet{B95} to the stars with $T_{\mathrm{eff}} 
< 4200 \ K$ and another relation \citet{BCP98} to the other cool PMS stars ($4200 \leq 
T_{\mathrm{eff}} \ [K] < 7000$). For the hotter stars ($T_{\mathrm{eff}} \geq 7000 \ K$), the 
color-effective temperature relations of \citet{SLB13} were used in the temperature estimation, 
because such intermediate-mass stars may have less vigorous accretion activities at the 
age of \objectname{NGC 2264}. The effective temperature of our sample is distributed in the 
range of 3500 $K$ to 20,000 $K$.

\subsection{Measurements of Equivalent Widths}

The main spectral feature in this study is the $^7$Li {\scriptsize \textsc{I}} $\lambda$6708 
resonance doublet. It is, in most cases, impossible to resolve the doublet separated by 0.1 \AA \ because of 
rotational broadening. The $^6$Li doublet also overlaps the red component 
of the $^7$Li {\scriptsize \textsc{I}} doublet. These isotopes are the primordial elements 
synthesized in the Big Bang. The standard Big Bang nucleosynthesis model predicts that 
only a small amount of $^6$Li ($\sim 10^{-4}$ of $^7$Li) was produced \citep{F11}. Indeed, 
the presence of $^6$Li has barely been confirmed in a few halo stars \citep{ALNPS06}. 
For the Sun, the fraction of $^6$Li to the total Li abundance is about 7.59\% \citep{AGSS09}. 
We assume that the contribution of $^6$Li to the observed absorption feature is negligible, 
and therefore the $^7$Li abundance is considered as the entire Li abundance. 

Each configuration of our observations comprises the spectra of different brightness PMS stars.  
The difference between the brightest and the faintest stars is about 3 mag in the $R$ band. For 
this reason, the spectra of the ten faintest stars were obtained with very low signal levels. In addition, 
the Li {\scriptsize \textsc{I}} $\lambda$6708 line was not detected in the spectrum of star S2587 (6300 $K$) 
due to the low signal-to-noise ratio although it is a relatively bright star ($V = 11.10$ mag). 
We could not find the Li {\scriptsize \textsc{I}} absorption line in the spectra of 19 bright PMS stars. 
The absence of the Li {\scriptsize \textsc{I}} line in the spectra of 16 of these stars 
may be attributed to the higher effective temperature ($T_{\mathrm{eff}} > 6500 \ K$). 
Only two PMS stars (the star W2350 and S2936 -- star identification used in \citealt{SBCKI08}) 
with a higher effective temperature exhibit a weak Li {\scriptsize \textsc{I}} line 
[$W_{\lambda6708}$(Li) $\sim$ 60 -- 80 m\AA]. We need to check the membership of the other 
three Li deficient stars. A detailed discussion on these stars will be made in Section 5. 

Gaussian profile fitting was used to measure the $W_{\lambda6708}$(Li) of 105 
PMS stars. The best fitting solution was obtained with the MPFIT packages \citep{M09}. 
The $W_{\lambda6708}$(Li) was computed by integrating the best-fit Gaussian profile. 
We also measured the upper and lower values of $W_{\lambda6708}$(Li) from the
 Gaussian profiles by altering the Gaussian width and amplitude according to the errors. 
The upper and lower values were adopted as the uncertainty of the $W_{\lambda6708}$(Li). 
The uncertainties of the $W_{\lambda6708}$(Li) for 16 faint PMS stars were too large 
($\mid \epsilon W_{\lambda6708}$(Li)$\mid$ $>$ 100 m\AA) for them to be used in the study. 
Figure~\ref{fig3} exhibits the spectra of 86 PMS stars with a detectable Li $\lambda$6708 
doublet. The best-fit Gaussian profiles are overplotted in the figure.

In Figure~\ref{fig4}, our $W_{\lambda6708}$(Li) measurements were compared with those 
of previous studies to check the external consistency. The mean difference between 
the $W_{\lambda6708}$(Li) of \citet{SKSJF99} and ours is $+7\pm44$ m\AA. There are 
three outliers in the comparison with those of \citet{THFHM15}. The small $W_{\lambda6708}$(Li) 
of the star S1731 and C22501 is related to the veiling effect of the Li {\scriptsize \textsc{I}} 
line probably due to the accretion activity at the time of observation (see Section 3.3). 
However, the spectrum of the other star S3061 was obtained with a reasonable signal, 
and does not show any accretion features, such as very strong H$\alpha$ and He 
{\scriptsize \textsc{I}} emission lines. No significant ultraviolet (UV) excess emission was 
also found from the photometry of \citet{SBL97}. The $W_{\lambda6708}$(Li) of the star 
may not be associated with the accretion activity. The difference may arise either from 
the measurement error in \citet{THFHM15} or from the intrinsic variation of the line strength, 
because the measurement of \citet{BLV16} for the star is in good agreement with ours 
($\Delta = +7.9$ m\AA). The mean difference between the $W_{\lambda6708}$(Li) 
of \citet{THFHM15} and ours is $+1\pm72 \ \mathrm{m\AA}$ if the outliers were excluded. 
Our results are also consistent with those of \citet{BLV16} with a mean difference of 
$-28\pm39$ m\AA \ for a wide range of $W_{\lambda6708}$(Li). Only one outlier 
(the star C34897) was found. A slightly large random error may be involved in the 
$W_{\lambda6708}$(Li) of the star because the star is somewhat faint ($V = 16.81$ mag). 
From these comparisons, we confirmed a good external consistency of $W_{\lambda6708}$(Li) 
measured in this study. 

The Li {\scriptsize \textsc{I}} $\lambda$6708 line is mainly blended with a neutral iron line 
located at 6707.4\AA. \citet{SJB93} presented an empirical relation between the 
$W_{\lambda6707.4}$(Fe {\scriptsize \textsc{I}}) and $B-V$ color obtained from old inactive stars 
without a Li {\scriptsize \textsc{I}} absorption line. The relation is expressed as $W_{\lambda6707.4}
(\mathrm{Fe \ {\scriptsize \textsc{I}}}) = 20(B-V) - 3 \ \mathrm{m\AA}$. If we use the 
dereddened $B-V$ colors, the $W_{\lambda6707.4}$(Fe {\scriptsize \textsc{I}}) would be underestimated 
by the UV excess of the $B-V$ color. Therefore, the expected photospheric $B-V$ colors 
were estimated from the dereddened $V-I$ colors using the $(B-V)_0$ versus $(V-I)_0$ relation 
for MS stars \citep{SLB13}. We corrected for the contribution of Fe {\scriptsize \textsc{I}} $\lambda$6707.4 
to the measured $W_{\lambda6708}$(Li) using the relation of \citet{SJB93}. 

Figure~\ref{fig5} shows $W_{\lambda6708}$(Li) as a function of effective temperature. From 
this relation, it naturally explains the reason why no Li was detected in the spectra of most 
intermediate-mass stars. An additional deblending was required for three PMS stars with 
significant line broadening (S1776, S2663, and S2715 -- red open circles in the figure). As seen 
in Figure~\ref{fig3}, Li {\scriptsize \textsc{I}} $\lambda$6708 and adjacent Fe {\scriptsize \textsc{I}} lines in the spectra 
of these stars are merged into a broad single line. According to the study of \citet{JJR16}, the stars have a high 
projected rotational velocity -- 133.6km s$^{-1}$ for S2663 and 193.8km s$^{-1}$ for S2715. High 
rotational velocity is likely responsible for the line broadening. Therefore, Fe {\scriptsize \textsc{I}} 
$\lambda$6705.1 and $\lambda$6710.3 can also affect the $W_{\lambda6708}$(Li) of these stars. 
In order to minimize their influences on $W_{\lambda6708}$(Li), 
we used the archival spectra of field stars with almost the same effective 
temperature as the stars, which were published by the Ultraviolet and Visual Echelle Spectrograph 
(UVES) Paranal Observatory Project (DDT Program ID 266.D-5655 -- \citealt{BJL03}). Since the 
UVES spectra were taken with a very high signal-to-noise ratio, their $W_{\lambda}$(Fe 
{\scriptsize \textsc{I}}) were directly measured by integrating the absorption lines after continuum 
normalization. A list of the field stars is presented with their correction values in Table~\ref{tab1}. 
The contribution of the Fe {\scriptsize \textsc{I}} lines to the $W_{\lambda6708}$(Li) was removed by 
subtracting the adopted $W_{\lambda}$(Fe {\scriptsize \textsc{I}}). Given the small error bars, several 
cool stars ($< 4000 \ K$) appear to actually have small $W_{\lambda6708}$(Li) at a given temperature. 

\subsection{The Effect of Veiling on the Li $\lambda$6708 Line}

The equivalent widths of spectral lines can be underestimated by excess continuum 
emission due to mass accretion activity \citep{CG98,LSKBK14}. We used the H$\alpha$ and He 
{\scriptsize \textsc{I}} $\lambda$6678 emission as a diagnostic tool to 
probe the veiling effect on $W_{\lambda6708}$(Li). The $W_{\lambda6563}$(H$\alpha$) and 
$W_{\lambda6678}$(He {\scriptsize \textsc{I}}) of cool PMS stars ($T_{\mathrm{eff}} < 5000 \ K$) 
were measured by integrating the observed emission lines. In the case of the He 
{\scriptsize \textsc{I}} line, we computed the equivalent width of the line core within a 
narrow wavelength range of 0.7\AA \ to minimize the contribution of the error in the 
continuum level. For this reason, the $W_{\lambda6678}$(He {\scriptsize \textsc{I}}) of the absorption line for 
moderately fast rotators may be underestimated, while strong (He {\scriptsize \textsc{I}}) 
emission stars may have slightly larger equivalent widths (smaller equivalent widths in absolute scale). It is known that the full width at 
10\% of peak H$\alpha$ flux (hereafter 10\% width) is closely related to the mass accretion rate 
of PMS stars \citep{WB03,MLBHC05}. The profiles of H$\alpha$ emission lines in the spectra 
of some active stars are depressed by either blueshifted or redshifted absorption components. 
The emission line profile 
can be restored by a Gaussian function fitted to the observed wing profile. The 10\% 
widths from the best-fit Gaussian profiles were then obtained for 66 stars. 

In the upper left-hand panel of Figure~\ref{fig6}, stars with a weak H$\alpha$ emission 
component tend to have relatively narrow 10\% widths (100 -- 250km$^{-1}$), while stars 
with a wide velocity width show a wide spread of $W_{\lambda6563}$(H$\alpha$). A similar 
feature can be found for the He {\scriptsize \textsc{I}} $\lambda$6678 line in the right-hand 
panel of the figure. Since the stars can be divided into two groups at 250 -- 270 km 
s$^{-1}$ in the distribution of the $W_{\lambda6563}$(H$\alpha$) and $W_{\lambda6678}$(He {\scriptsize 
\textsc{I}}), we adopted 270 km s$^{-1}$ as a selection criterion for the accreting stars. This 
value is the same criterion as that of \citet{WB03}, but larger than that of \citet{MLBHC05}. 
In addition, stars showing a He {\scriptsize \textsc{I}} $\lambda$6678 emission line were also 
selected as accreting stars because the He emission line is known to form in postshock 
regions \citep{BEK01}. It is confirmed that a total of 27 stars have either a broad H$\alpha$ 
emission line or a He {\scriptsize \textsc{I}} emission line, or both. If the spectral lines are veiled 
by accretion activities, $W_{\lambda6708}$(Li) may be underestimated.
The lower panels of Figure~\ref{fig6} show the variations of $W_{\lambda6708}$(Li) with respect 
to $W_{\lambda6563}$(H$\alpha$) and $W_{\lambda6678}$(He {\scriptsize \textsc{I}}). Stars with strong 
emission lines shows a tendency to have a smaller $W_{\lambda6708}$(Li). 

The highly veiled and little veiled stars may have different surface coverage of the shocked 
regions \citep{CG98}, and the lifetime of hot spots created by the accreted materials may 
also be different for a different accretion rate. The intensive time series observations of 
some PMS stars in \objectname{NGC 2264} show aperiodic or periodic brightening 
events in their light curves on a timescale of several hours to 30 days \citep{SCB14,SCR16}. 
The aperiodic events are related to the variable mass accretion rate, and the rotational modulation 
of long-lived hot spots is responsible for the periodic light curves. These stars tend to have 
bluer $u-g$ than other stars at a given $g-r$ (i.e. UV excess emission), and a strong 
H$\alpha$ emission line. The variation of $W_{\lambda6708}$(Li) of a few accreting stars was 
monitored with the continuum level in one of the previous studies. $W_{\lambda6708}$(Li) appears 
to weaken as the star brightens. This fact may indicate that the stars with small 
$W_{\lambda6708}$(Li) ($\lesssim$ 300 m\AA) at $\sim 4000 \ K$ in Figure~\ref{fig5} are highly 
veiled stars.

\subsection{Lithium Abundance}

\citet{SJB93} published curves of growth (COG) for the Li {\scriptsize \textsc{I}} $\lambda$6708 
resonance doublet under the conditions of local thermodynamic equilibrium (LTE) for the 
effective temperature range of 4000 $K$ -- 6500 $K$. The deblended $W_{\lambda6708}$(Li) 
of the PMS stars hotter than 4500 $K$ was converted to LTE abundances using these COG. 
The LTE abundance of the cooler stars was determined 
from the COG of \citet{ZBP02}. The abundance is based on the scale of $\log N(\mathrm{H}) = 12$. 
We also corrected for non-LTE effects using the calculations of \citet{CRBS94} for stars hotter than 
$4500 \ K$. The correction values are about 0.1 -- 0.2 dex for $A(\mathrm{Li})_{\mathrm{LTE}} = 3.0$ 
the solar metallicity. As \citet{PRMG95} has shown that the difference between the LTE and non-LTE 
abundances becomes negligible at $T_{\mathrm{eff}} < 4500$K, the non-LTE effects were not taken 
into account for the cooler stars. We summarize all the measurements in Table~\ref{tab2}.

Figure~\ref{fig7} displays the Li abundance of the PMS members in \objectname{NGC 2264} 
with respect to the effective temperature. We investigated the distribution of the Li abundance 
as shown in the left-hand panel of Figure~\ref{fig8}, where only stars hotter than 4300 $K$ were 
used to avoid any variation of Li abundance with age. The abundance peak appears at 3.2, and a
1$\sigma$ error is about 0.2 dex. The result is consistent with the initial Li abundance in nearby 
star forming regions \citep{MRP92,CSL95,K98}, as well as the remaining Li abundance of solar system
meteorites \citep{AGSS09}. It may be the initial abundance of the natal cloud from which 
the cluster formed. 

We present several PMS isochrones from two different evolutionary model grids in 
Figure~\ref{fig7} \citep{SDF00,BHAC15}, where the initial Li abundance of 3.2 was adopted. 
The Li abundance of the Sun [A(Li) = 1.05 -- \citealt{AGSS09}] is also shown as the 
solar symbol with the 4.6 Gyr isochrone from the models in order to check which 
model properly predicts the solar Li abundance. However, neither of the models can explain the solar 
Li abundance. A possible explanation for the low Li abundance of the Sun may be
non-standard depletion processes, such as rotational mixing and microscopic diffusion 
during the MS stage \citep{CDP95} or the imperfect treatment of the convection zone in the PMS 
models.

The two evolutionary model grids predict a different amount of Li at the same age for cool stars 
($\sim 4000 \ K$). In the Siess models, Li is depleted by 0.7 dex at 5 Myr and 2.5 dex at 7 Myr, 
and little Li remains at 10 Myr. On the other hand, the Baraffe models predict a rather smaller 
depletion of Li, i.e. about 0.3, 0.7, and 1.7 dex at 5, 7, and 10 Myr, respectively. Our measurements 
show a wide span of Li abundance for the cool stars [$A\mathrm{(Li)} = -0.8$ to $3.9$]. However, the stars 
pretending to be old are all the highly veiled stars. They show a small Li abundance because of their 
underestimated $W_{\lambda6708}$(Li). Had we included these stars in the relative age estimation, 
the age spread of \objectname{NGC 2264} would be larger than 10 Myr. This is a similar 
situation to the photometric age estimation from the CMDs, where the CMDs are a kind 
of snapshot taken at an arbitrary epoch, and it is very difficult to discriminate the displacement of colors 
and magnitude due to the veiling effect, as well as due to the obscuration of clumpy 
material. On the other hand, the veiling effect on the Li abundance can be recognized by using the 
correlations between the strength of various emission lines and the $W_{\lambda6708}$(Li) as above. 
The obscuration by disk material may only be a minor source affecting the measured $W_{\lambda6708}$(Li) 
\citep{SCB14}. The presence of a large age spread among cool PMS stars can be 
confirmed from the distribution of their Li abundance if they are genuine Li depleted old PMS members 
in the cluster. If we exclude the highly veiled stars from our sample, the apparent age spread 
among the PMS stars is likely about 5 -- 7 Myr. But given the size of the observational errors, 
the age spread may in fact be smaller than this value.
 
\section{AGE SPREAD AND CLUSTER FORMATION TIMESCALE}
The age of PMS stars can be estimated from the A(Li)-$T_{\mathrm{eff}}$ plane using the 
evolutionary models, however there are several practical difficulties. It is impossible to estimate the 
age of PMS stars with A(Li) larger than 3.2, and Li depletion is insignificant within the first 3 Myr. 
Somewhat large errors in age estimation could be introduced by the scatter of Li abundance. 
The observed scatter may be a mix of observational errors due to their faintness, intrinsic 
variations due to stellar activity, and evolutionary effects. The variation of Li abundance due to 
stellar activity are likely much smaller than the observational error \citep{PCRG93,M98,SSW15}, 
therefore the effects of stellar activity on Li abundance can be ignored. For higher mass stars 
($> 1M_{\sun}$), Li depletion is terminated on a very short timescale because their radiative core 
is rapidly developed \citep{SHJMN14}. The Li abundance of the higher mass stars may be insensitive 
to effects of evolution within 10 Myr. Hence, observational errors can be exclusively inferred 
from the distribution of their Li abundance, as shown in the left-hand panel of Figure~\ref{fig8}. If 
the error from the A(Li) distribution is a true uncertainty due to observational error over the whole 
mass range, an observed scatter that is larger than the observational error may indicate the underlying 
age spread. Instead of a direct age estimation of individual stars, we conducted multiple sets of 
simulations for a range of observational errors in order to estimate the underlying age spread.

Stars in the effective temperature range of 3800 $K$ to 4300 $K$ were selected as a subsample 
because the Li abundances of these stars are most sensitive to age. Highly veiled stars 
[$W_{\lambda6708}$(Li) $<$ 300 m\AA] were excluded from the subsample of 41 PMS stars. 
A total of seven model clusters were created to reproduce the distribution of Li abundances using 
a Monte-Carlo method. The individual cluster contained 1000 stars in the mass range  0.4 $M_{\sun}$ to 
1.4 $M_{\sun}$. The underlying initial mass function was assumed to be $\Gamma = -1.7$ 
($\alpha = 2.7$ -- \citealt{SB10}). The age of the model clusters was set to 3 Myr \citep{SB10}, 
and age spreads of  0 -- 6 Myr were applied to the clusters with a 1 Myr interval, respectively. 
The evolutionary models of \citet{BHAC15} for PMS stars were adopted for assigning the 
effective temperature and Li abundance of individual stars of the model clusters. 
The observational error adopted had the same distribution as our measurements 
(1$\sigma$ = 0.2 dex, see the left-hand panel of Figure~\ref{fig8}). 

The middle-panel of Figure~\ref{fig8} exhibits the cumulative distribution of Li abundance 
for the subsample and model clusters over the same effective temperature range. The extent 
of the tail toward small Li abundance becomes evident for a large age spread. The Li 
abundance distribution of the model cluster (red solid line) with an age spread of 4 Myr 
seems to well reproduce the observed one (triangles). The abundance distributions of 
the seven model clusters were assumed to be the same as the parent population for a given age spread. We 
quantitatively investigated which parent population gives the best match to the observed Li 
distribution using the Kolmogorov--Smirnov test. The probabilities in percentile are 
displayed in the right-hand panel of Figure~\ref{fig8}. The 4 Myr spread model shows a 96.5\% 
match to the Li distribution of the subsample. The 3 Myr and 5 Myr spread models have a similar 
distribution at $\sim 70$\% confidence level. 

We performed the same simulations for the PMS evolutionary models of \citet{SDF00}. 
The results from these simulations are also plotted by open circles in the right-hand panel of 
Figure~\ref{fig8}. The trends are very similar to each other, however the age spread of the 
model cluster which most closely reproduces the observed distribution of Li abundances was 3 Myr, 
1 Myr smaller than the result based on the Baraffe models. The age spread of other probable 
model clusters at a confidence level greater than 70\% appears at 2 Myr and 4 Myr, respectively. Our 
results indicate that the extended star formation scenario ($\Delta \tau \geq 10$ Myr) can be 
ruled out for \objectname{NGC 2264} regardless of evolutionary models. The formation of 
optically visible stars in the cluster may be complete within 3 -- 4 Myr. 

On the other hand, multiple sets of the same simulations were also conducted with different observational 
errors of 1$\sigma$ = 0.05, 0.10, 0.15 and 0.30 dex in order to test the validity of the adopted 
observational error. These model clusters give different cumulative distributions in Li abundance. The slope of the 
cumulative distributions of the model clusters adopting an error of 0.05 dex is steeper than that of the 
observed subsample, while the model clusters with a large error of 0.3 dex give shallower slopes. We investigated 
the similarity between the subsample and the model clusters through the Kolmogorov--Smirnov test. 
Table\ref{tab3} shows the underlying age spreads of the most probable model cluster and confidence 
levels. While the model clusters adopting the smallest and largest errors give completely different 
age spreads with low confidence levels, the age spread of the others are very similar 
to each other. The confidence level for the observational error adopted in this study (0.2 dex) 
appears higher than the others. Hence, the adopted error of 0.2 dex is likely the true uncertainty. 

According to \citet{FHS06} and \citet{SSB09}, \objectname{NGC 2264} is comprised 
of at least three subclusters -- \objectname{S Mon}, \objectname{Spokes}, and \objectname{Cone} 
nebula regions as shown in Figure~\ref{fig9}. In addition, a low stellar density region (halo) surrounds these 
subclusters \citep{SBCKI08}. The largest subcluster \objectname{S Mon} contains not only the high-mass 
multiple system S Mon (O7V + early B + ? -- \citealt{SMW11}) but also a large 
number of Class II objects. However, the number of Class I objects is much 
smaller than that of the other embedded or partially embedded subclusters, the
\objectname{Spokes} and \objectname{Cone} nebula regions. It may imply the 
presence of an age difference among these subclusters. \citet{SB10} suggested an outside-in 
sequential star formation scenario within \objectname{NGC 2264} 
from the age distribution of the PMS stars. A nearby supernova explosion about 6 -- 7 
Myr ago was suspected as the source of the external trigger. 

The presence of the subclusters can also be found in the velocity field obtained from 
$^{13}$CO observations and stellar radial velocity survey \citep{FHS06,THFHM15}. The radial 
velocity distribution of the cluster members shows a velocity gradient of 10 km s$^{-1}$ along 
the north-south direction. The stars in the \objectname{S Mon}, \objectname{Spokes}, and 
\objectname{Cone nebula} regions are moving away from the Sun by $V_{\mathrm{helio}} 
= $19 -- 24 km s$^{-1}$, 18 -- 21 km s$^{-1}$, and 20 -- 24 km s$^{-1}$, respectively. 
$^{13}$CO gas shows almost the same kinematics as that of the member stars. Based on 
the simulations of \citet{BH04}, the velocity structure of the subclusters was interpreted as 
the result of gravitational focusing in an inhomogeneous molecular cloud with a finite size and 
non-spherical shape.

Most of stars in the subsample are located either in the \objectname{S Mon} region or 
 the halo region (see the right-hand panel of Figure~\ref{fig9}). If the star formation 
history suggested by \citet{SB10} is true, an age spread of 3 -- 4 Myr may be the 
time duration of the star formation activity propagated from the halo to the \objectname{S Mon} region. 
A large number of stars in the \objectname{Spokes} and \objectname{Cone} nebula regions are still embedded in the 
cloud. The high number ratio of Class I to Class II objects 
indicates that the subclusters in the regions have been formed recently relative to the 
other regions. If we adopt the median age of the individual subclusters from Table 2 of 
\citet{SB10}, the age difference between the halo and \objectname{Spokes} + 
\objectname{Cone nebula} regions is about 1 Myr. Hence, the stellar population in the 
entire region of  \objectname{NGC 2264} may have formed within 5 Myr.  

\section{DISCUSSIONS ON THE Li DEFICIENT STARS}
Figure~\ref{fig10} exhibits the ($V, V-I$) CMD of the PMS stars observed in this work. 
It is difficult to identify the Li {\scriptsize \textsc{I}} $\lambda$6708 doublet in the 
spectra of PMS stars (blue square) with blue $V-I$ colors ($\lesssim 0.5$) because 
these stars, in general, show a large rotational broadening and an intrinsically small 
$W_{\lambda6708}$(Li) due to their higher effective temperature. Actively accreting 
PMS stars (open triangle) are mostly subsolar mass stars. These stars show a 
large spread in $V$ magnitude at a given color, and this fact shows that accretion 
activity is a source of the observed spread. On the other hand, the faint end of the 
observing selection for bright stars appears at $V \sim 13.5$ mag, and the limiting magnitude 
of the faint star selection is about $V = 17$ mag. $W_{\lambda6708}$(Li) could be 
barely measured for some of the faint stars near the limiting magnitude, but the 
uncertainty is somewhat large (asterisks). For the other stars in the faintest group, 
we could not find any detectable Li line in their spectra because of low signal-to-noise 
ratios. Further observations with a larger telescope for these stars are required 
to identify the presence of the Li line.

There are three Li deficient stars among relatively bright targets. The properties 
of these stars were investigated using available information in the previous studies. 
We list the stars with photometric data in Table~\ref{tab4} and plot their spectra 
in Figure~\ref{fig11}. The star S2223 (\objectname{V642 Mon}) was selected as an X-ray 
emission member of \objectname{NGC 2264} from the {\it Chandra} X-ray data \citep{SBC04}. 
The star shows a weak H$\alpha$ absorption line and broad metallic lines. \citet{FHS06} 
identified the star as a spectroscopic binary from their multi-epoch 
spectroscopic observations. The star is brighter than other stars at a given color in 
Figure~\ref{fig10}. If we assume that the object is a nearly equal-mass binary system, 
the brightness of individual component will be lower by about 0.75 mag. However, each 
component is still brighter than other PMS members with almost the same colors. We 
considered the object as a foreground active binary system to reconcile the two independent 
facts, their high brightness and Li deficiency.

A pair of late-type stars is likely to constitute the binary system given the weak H$\alpha$ 
absorption line and the red $V-I$ color. Since the foreground reddening toward 
\objectname{NGC 2264} is nearly zero, the spectral type of the primary star could be K2 
according to its color. We considered a binary system comprising two equal-mass 
MS stars. The absolute magnitude of the equal-mass binary is 5.5 mag where the absolute 
magnitude of a K2V star is about $M_V =  6.3$ mag \citep{SLB13}. Using the $V$ magnitude of the object in Table~\ref{tab4}, 
the distance modulus of the binary system is about 6.3  mag ($d = 182$ pc). On the other hand, 
a binary system with a very small mass ratio (e.g. K2V + M3V) was also considered. In this 
case, the composite absolute magnitude is almost the same as the absolute magnitude of a 
single K2V star ($M_V = 6.3$ mag -- \citealt{SLB13}). The distance to this binary system is then 
about 124 pc. Given that this object was identified as a double-lined binary by \citet{FHS06}, 
the mass of the companion star must be comparable to that of the K2V star. The former case 
is a more appropriate situation to explain the overall properties of the object. Hence, S2223 
is likely an X-ray emitting binary system consisting of two early K-type stars within 182 pc.

The star S3755 is also known as an X-ray emission star \citep{DSPP07}. In Figure~\ref{fig10}, 
the star appears brighter than other stars at a given color. Therefore, the Li deficiency could be 
explained by other factors. The high resolution UVES spectrum of \objectname{HD 320868} 
(K5 -- \citealt{BJL03}) was used as a template to investigate its cross correlation function. 
In Figure~\ref{fig12}, the cross-correlation function shows two components, a sharp peak on 
a broad profile. The star seems to have a composite spectrum of two stars. The heliocentric 
radial velocities of the sharp and broad components are about 85.5 and 61.4 km s$^{-1}$, 
respectively. The narrow component including H$\alpha$ absorption line in the spectrum of 
S3755 shows a reasonable match to that of \objectname{HD 320868} with a cross correlation 
amplitude of 0.59. In addition, the red $V-I$ color indicates that one of the components may 
be a late-K-type star. Although the cross correlation functions with the UVES spectra of 
other G8V and K2V stars were also investigated to constrain the spectral type of the 
companion, the cross correlation amplitude insignificantly varied in the spectral type range. A 
appreciable contribution of the K-type star to the spectrum of S3755 may imply that the other 
companion star may have a comparable flux.

If the binary system had comprised two young K-type stars, the Li {\scriptsize \textsc{I}} 
line would be detectable. The spectrum of S3755 shows an H$\alpha$ line in absorption, 
while most of the K-type members of \objectname{NGC 2264} in this study show an H$\alpha$ 
emission line. We, therefore, considered the object as a pair of two K5V stars in the foreground 
based on the $M_V$-spectral type-color relations of \citet{SLB13} as applied to S2223. 
The absolute magnitude of the binary system was estimated to be 6.5 mag, where an absolute 
magnitude of 7.3 mag was adopted for each star. If the foreground reddening toward the stars 
is assumed to be zero, the distance modulus of the binary system is about 7.4 mag, equivalent 
to 302 pc. The same calculation was also carried out for a pair of K0V and K5V stars. In this 
case, the object would be located at 442 pc from the Sun. The photometric and spectroscopic 
properties of S3755 could be explained by those of a foreground binary system consisting 
of late-type stars with an X-ray emission within 300 -- 500 pc.

The star W2320 (\objectname{LkH$\alpha$ 4}) was selected as the member of the 
cluster from H$\alpha$ photometry \citep{SBCKI08}. The spectrum of the star shows a 
moderately strong H$\alpha$ emission line with a blueshifted absorption component 
(Figure~\ref{fig11}). The 10\% width of the emission line is about 529 km s$^{-1}$. The 
most distinctive feature is the presence of the diffuse interstellar band (DIB) at 6614\AA. Such 
an interstellar absorption line could not be found in the spectra of other members. The 
$W_{\lambda6614}$(DIB) is 295$^{+31}_{-29}$ m\AA, and it implies that the reddening toward the star 
is about $E(B-V) = 1.35$ from the $W_{\lambda6614}$(DIB) versus $E(B-V)$ relation \citep{LMZ15}. 
Since this star is located toward the less reddened field region of \objectname{NGC 2264}, 
the reddening of the stars is unlikely to be caused by the molecular cloud 
associated with NGC 2264. We speculate that the star may be a background cataclysmic 
variable given its spectral features are similar to those of a few cataclysmic variables, such 
as \objectname{V363 Aur} or \objectname{SDSSJ143317.78+101123.3} \citep{HLG82,TRD09}. 
 
\section{SUMMARY AND CONCLUSIONS}
The duration of the star formation activity in a cluster is basic information to understand 
the formation process of star clusters. A large number of previous studies have
attempted to estimate the cluster formation timescale from the CMDs using various evolutionary 
models for PMS stars. Their age spread was found to be typically larger than 5 Myr, up to 20 Myr 
for extreme cases \citep{CSv15}. However, a few critical limitations of the age estimation 
from the evolutionary models have been pointed out \citep{SBL97,H03}, e.g., systematically old age for 
intermediate-mass PMS stars. The observational uncertainties are too large to precisely 
estimate the age of individual PMS stars relative to a luminosity interval among different age 
isochrones. In this current study, Li abundance was used as a relative age indicator among PMS 
stars, as an alternative way to investigate the age spread in a cluster.  \objectname{NGC 2264} is an ideal 
target for this study because (1) the cluster is very close, (2) more than one thousand members 
are known, (3) the foreground reddening is nearly zero, (4) the differential reddening across 
the cluster is very small \citep{SBL97}, and (5) a large age spread ($\sim$10 Myr) has been 
reported in previous studies \citep{PaSt00}. We summarize our results and conclusion on the 
age spread of the PMS members in \objectname{NGC 2264} as below.

A total of 134 PMS members were observed with the multi-object echelle spectrograph 
Hectochelle attached to the 6.5m MMT. Although the Li {\scriptsize \textsc{I}} $\lambda$6708 
resonance doublet could not be identified in the spectra of most intermediate-mass PMS 
stars, we successfully detected the Li line in the spectra of 86 PMS stars and 
measured $W_{\lambda6708}$(Li) using Gaussian profile fitting. The contributions 
of adjacent neutral iron lines to the $W_{\lambda6708}$(Li) were removed using 
the empirical relation of \citet{SJB93} and UVES archival spectra of field stars 
\citep{BJL03} for the blended case. 

The $W_{\lambda6708}$(Li) of cool stars ($3600 < T_{\mathrm{eff}} [\mathrm{K}] < 4300$) shows a 
large spread of 400 m\AA. We investigated a few relations between the $W_{\lambda6708}$(Li) and 
indicators of accretion activity, such as H$\alpha$ and He {\scriptsize \textsc{I}} $\lambda$6678 
emission lines. The relations indicated that the $W_{\lambda6708}$(Li) can be affected by the 
veiling effect for the stars with 10\% width of a H$\alpha$ emission line $>$ 270 km s$^{-1}$ or 
a He {\scriptsize \textsc{I}} $\lambda$6678 emission line. Cool PMS stars ($T_{\mathrm{eff}} < 4500 \ K$) 
with $W_{\lambda6708}$(Li) smaller than 300 m\AA \ turned out to be highly veiled stars, 
not stars with weak Li lines due to their older age.

LTE Li abundance was derived by interpolating the Li {\scriptsize \textsc{I}} $\lambda$6708 COGs 
of \citet{SJB93} and \citet{ZBP02}, and then corrected for non-LTE effects using 
the relations of \citet{CRBS94}. The initial Li abundance of the natal cloud 
which forms \objectname{NGC 2264}, was estimated to be A(Li)$= 3.2 \pm 0.2$ from the 
abundance distribution of the stars with the effective temperature of 4300 $K$ to 6500 $K$. 
The result is in good agreement with that of previous studies for nearby star forming regions. 

The age of individual PMS stars could not be estimated from Li isochrones because of the large 
observational errors relative to the variation of Li abundance with age. Instead, the underlying 
age spread among PMS stars ($3800 < T_{\mathrm{eff}} \ [K] < 4300$), except for the highly 
veiled stars, was inferred from multiple sets of Monte-Carlo simulations considering the observational error. 
The simulations adopting an age spread of 3 -- 4 Myr appear to well reproduce the distribution 
of observed Li abundances. As most of the stars used in this study are located in \objectname{S Mon} region 
or halo region, and given the embedded population in the \objectname{Spokes} and \objectname{Cone} 
nebula regions, the formation timescale of the entire region of \objectname{NGC 2264} is 
likely to be about 4 -- 5 Myr. The extremely extended star formation scenarios 
\citep{H62,PaSt99,PaSt00,PaSt02} can be ruled out, at least for \objectname{NGC 2264}.

We also made careful discussions of three Li deficient cool stars. The star ID S2223 was known 
to be a spectroscopic binary \citep{FHS06}. The star S3755 also turned out to be a double-lined 
binary from cross correlation techniques. The photometric and spectroscopic properties indicate 
that both these objects are foreground binary systems consisting of two late-type stars. 
In the case of W2320, the object shows a strong interstellar absorption line at 
6614 \AA \ in its spectrum. This object is therefore considered a background star from the large 
extinction obtained from $W_{\lambda6614}$(DIB). The spectral features of W2320 resemble  
those of cataclysmic variables. These Li deficient stars are unlikely to be members of \objectname{NGC 2264}.

\acknowledgments

The authors thank the anonymous referee for useful comments and suggestions. The 
authors would also like to thank Perry Berlind, Mike Calkins, and Nelson Caldwell at SAO for assisting 
with Hectochelle observations. This work has used the data obtained under the K-GMT Science 
Program (PID: [15A-MMT-001 and 15B-MMT-005]) funded through Korean GMT Project operated 
by Korea Astronomy and Space Science Institute (KASI) and partly supported by KASI grant 
2016183201. HS acknowledge the support of the National Research Foundation of Korea (NRF) 
funded by the Korea Government (MOE, grant No 2015058444).

{\it Facilities:} \facility{MMT (Hectochelle)}.

\clearpage


\begin{figure}
\epsscale{.80}
\plotone{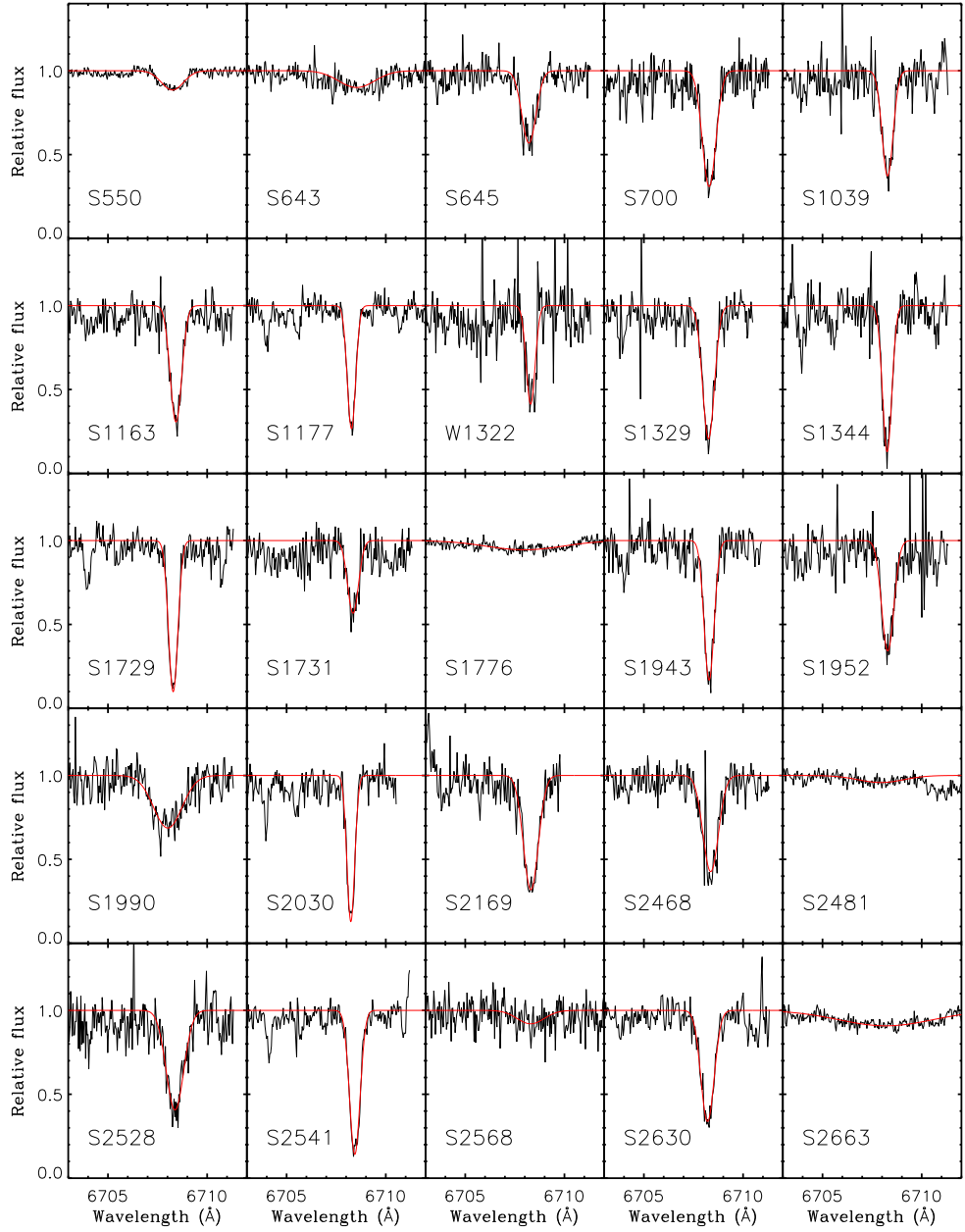}
\caption{Spectra of 86 pre-main sequence stars with a detectable Li $\lambda$6708 resonance doublet. The superimposed red solid curves 
represent the best-fit Gaussian profile for a given star. Object name is labeled at the bottom of each panel.}\label{fig3}
\end{figure}

\begin{figure}\figurenum{3}
\epsscale{0.80}
\plotone{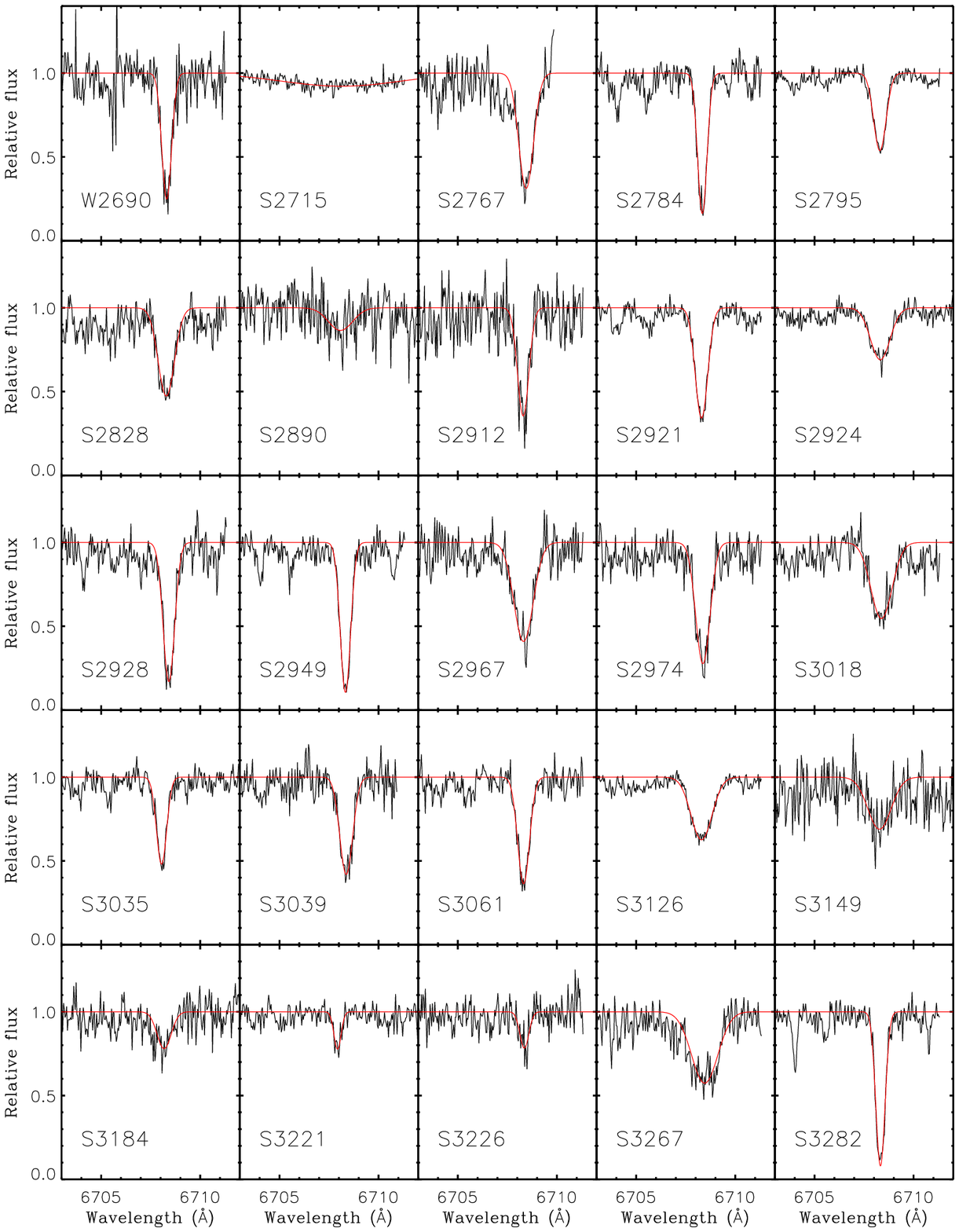}
\caption{Continued}
\end{figure}

\begin{figure}\figurenum{3}
\epsscale{0.80}
\plotone{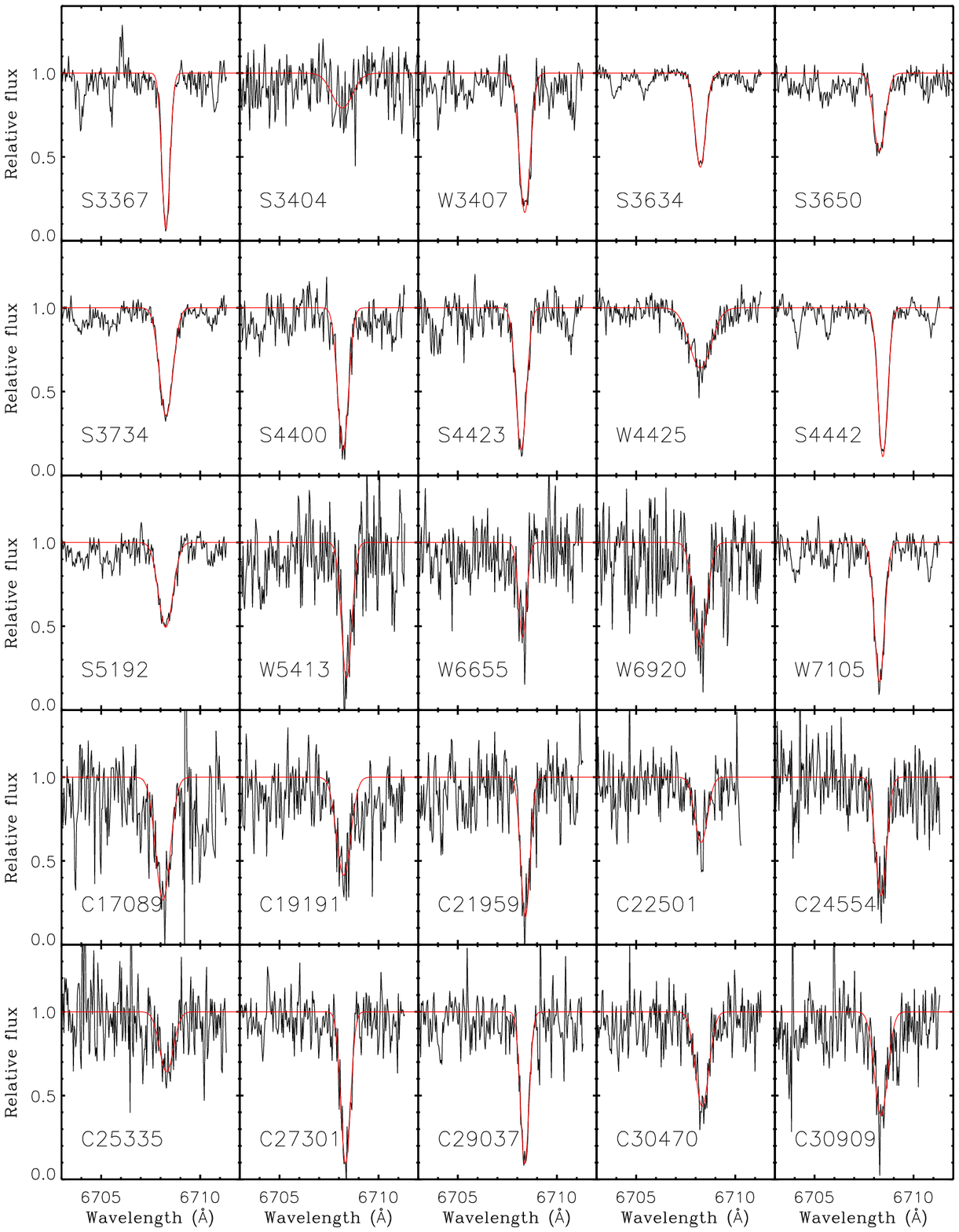}
\caption{Continued}
\end{figure}

\begin{figure}\figurenum{3}
\epsscale{0.80}
\plotone{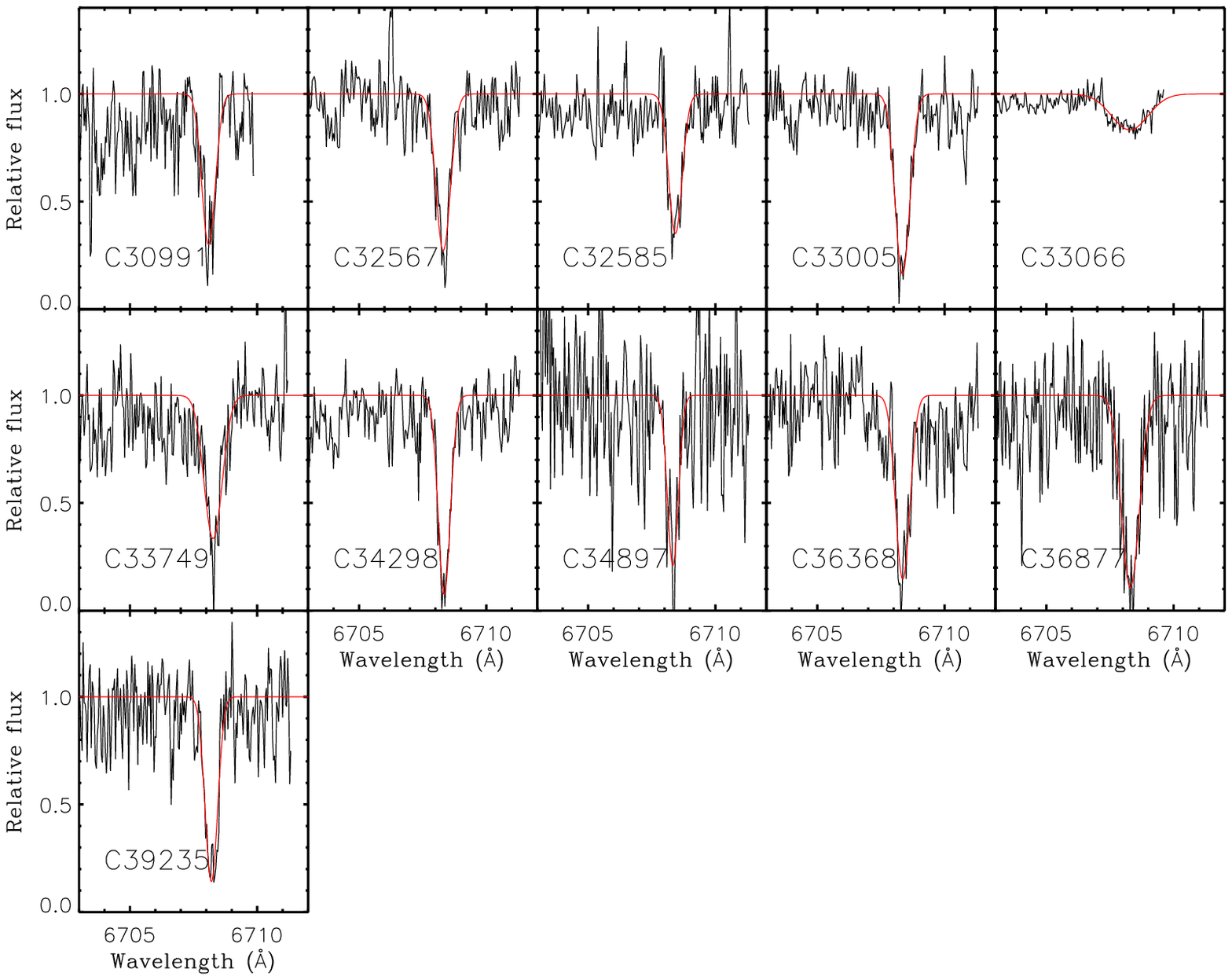}
\caption{Continued}
\end{figure}

\begin{figure}
\epsscale{.80}
\plotone{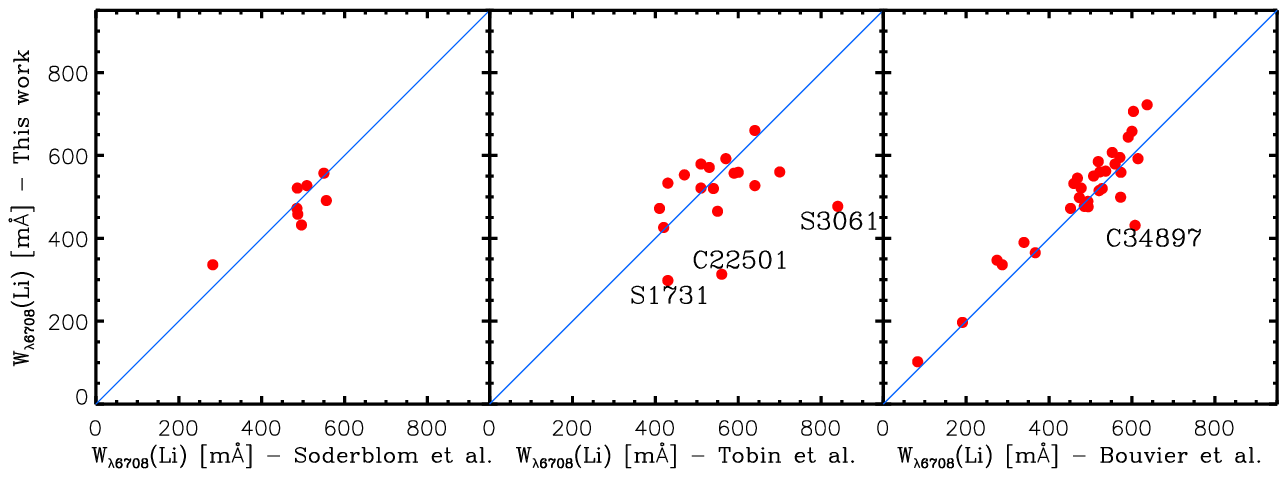}
\caption{Comparisons of the Li equivalent width measured in this work with that in other previous 
studies (left -- \citealt{SKSJF99}; middle -- \citealt{THFHM15}; right -- \citealt{BLV16}). The mean 
differences ($\equiv$ $\langle$ other $-$ this $\rangle$) are $+7 \pm 44 $, $+1\pm72$, and $-28\pm39$ 
m\AA \ respectively. The star ID of outliers is marked in each panel.}\label{fig4}
\end{figure}

\begin{figure}
\epsscale{.80}
\plotone{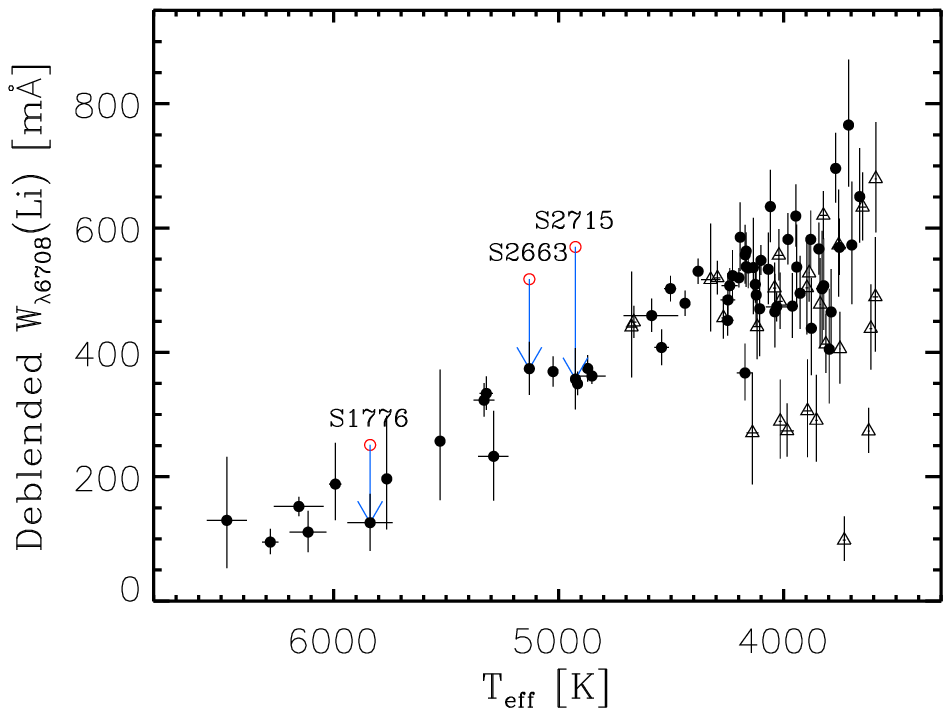}
\caption{Deblended equivalent widths of Li {\scriptsize \textsc{I}} $\lambda$6708 resonance doublet 
for 86 pre-main sequence members. Open circles denote the equivalent widths of four rapidly 
rotating stars, whose Li {\scriptsize \textsc{I}} $\lambda$6708 doublet is blended with 
Fe {\scriptsize \textsc{I}} $\lambda$6705.1 and $\lambda$6710.3 absorption lines. Arrows indicate 
the deblended values of the stars. Stars showing a He {\scriptsize \textsc{I}} $\lambda$6678 
emission line or the broad H$\alpha$ emission ($>$ 270 km s$^{-1}$) are marked by open triangles. }\label{fig5}
\end{figure}

\begin{figure}
\epsscale{.90}
\plotone{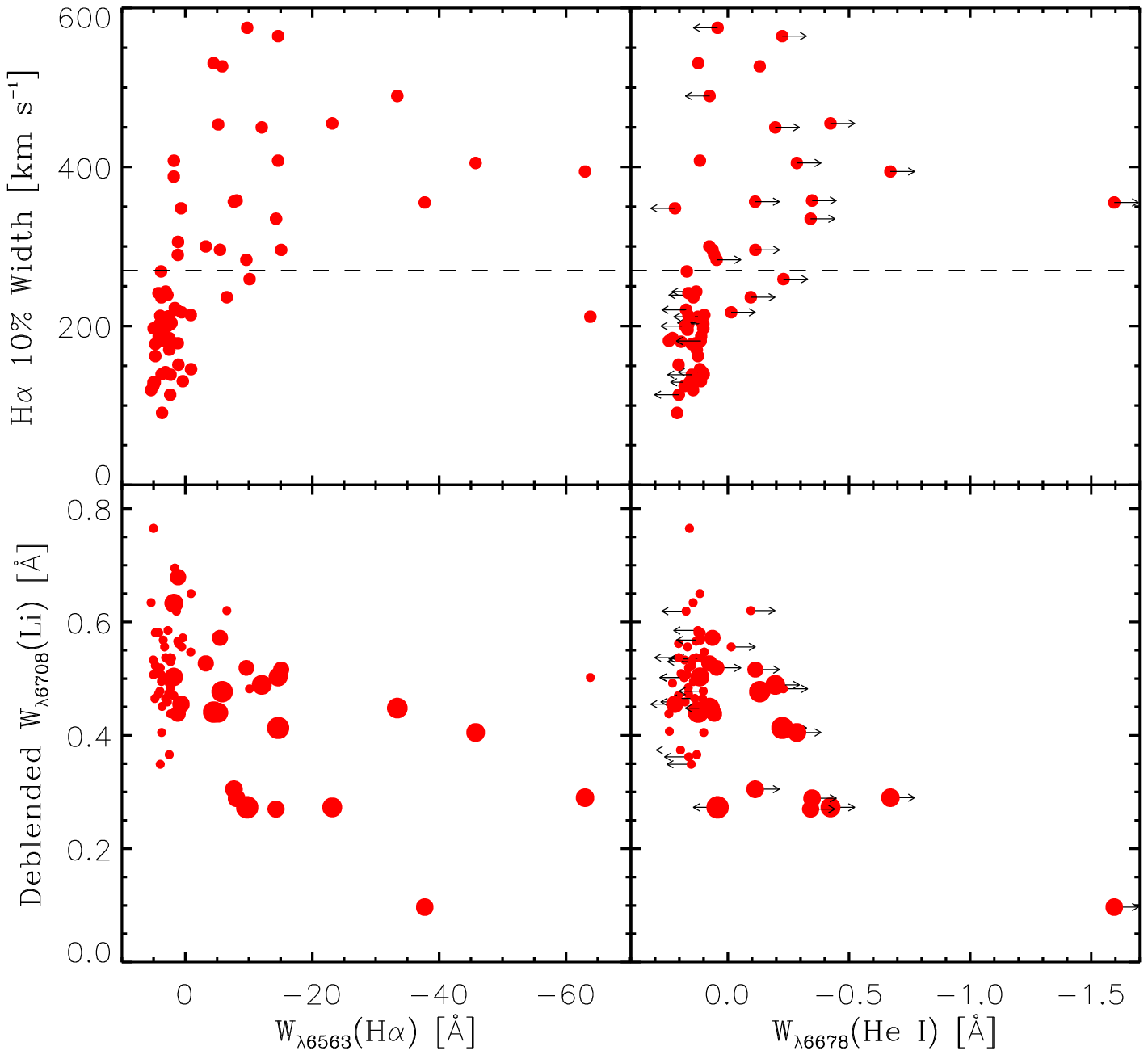}
\caption{Spectral features associated with accretion activities. Upper left and upper right panels show 
full widths at 10\% of the peak H$\alpha$ flux with respect to the equivalent widths of a H$\alpha$ 
and He {\scriptsize \textsc{I}} $\lambda$6678 emission lines, respectively. The dashed line represents 
the selection criterion for actively accreting stars suggested by \citet{WB03}. Lower left and lower right panels 
exhibit the variation of the equivalent widths of the Li {\scriptsize \textsc{I}} line against those of H$\alpha$ and 
He {\scriptsize \textsc{I}} $\lambda$6678 emission lines. Stars with broad 10\% widths of H$\alpha$ 
emission lines ($>$ 270 km s$^{-1}$) are marked by large symbols. The arrows denote 
upper and lower limits of equivalent widths of the He {\scriptsize \textsc{I}} $\lambda$6678 emission 
line.  }\label{fig6}
\end{figure}

\begin{figure}
\epsscale{.90}
\plotone{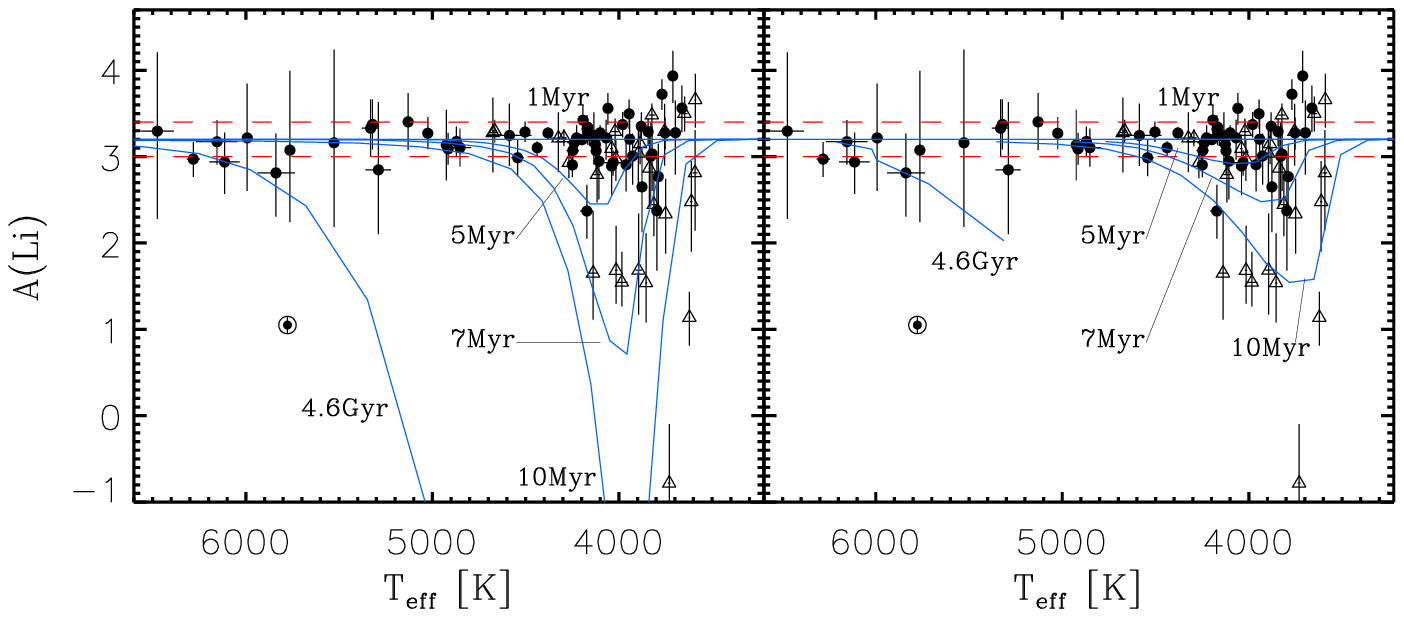}
\caption{Li abundance of the pre-main sequence (PMS) members in NGC 2264. Solid lines represent 
the 1, 5, 7, 10 Myr and 4.6 Gyr isochrones from two evolutionary models for PMS 
stars \citep{SDF00,BHAC15}. The region enclosed by the two dashed lines exhibits the range of the 
initial Li abundance $\mathrm{A(Li)} = 3.2 \pm 0.2$. For comparison, the Li abundance of the Sun 
is presented by the solar symbol. Stars showing accretion activities are marked by open triangles. }\label{fig7}
\end{figure}

\begin{figure}
\epsscale{1.00}
\plotone{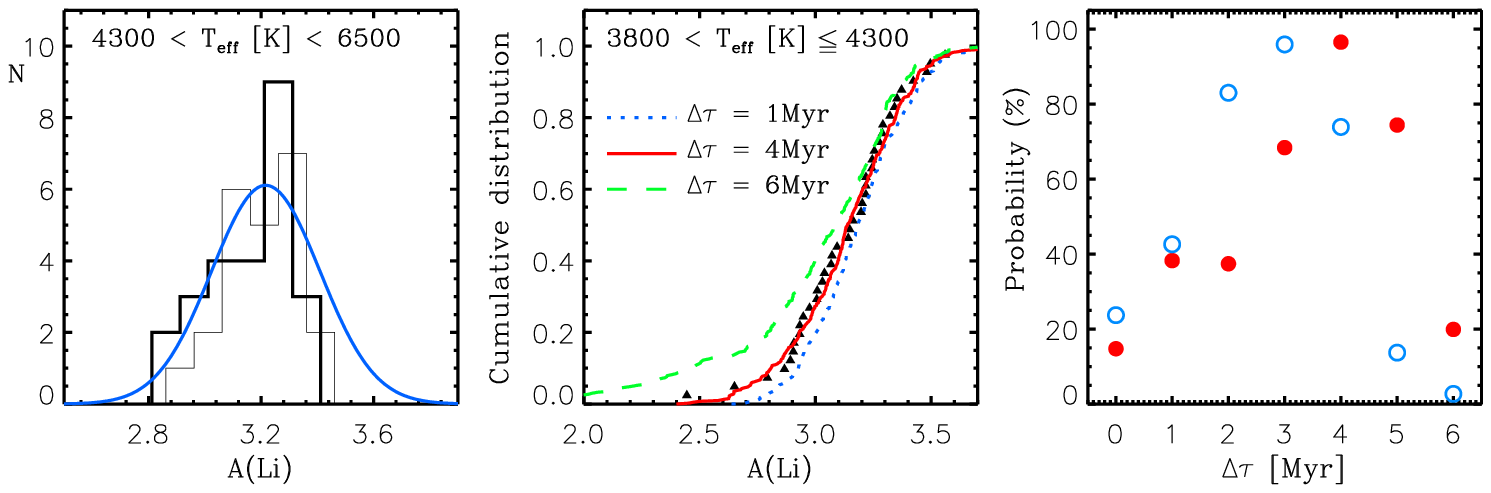}
\caption{Distribution of the observed Li abundance of hot PMS stars (left), the cumulative distribution of 
Li abundances of cool PMS stars and comparison of the observed Li abundance with that of model clusters 
(middle), and probability distribution of the age spread of NGC 2264 (right). Thick solid histogram in the 
left-hand panel was obtained from stars with effective temperature of 4300 $K$ -- 6500 $K$ to avoid evolutionary effects on the Li abundance for cooler stars. Thin solid histogram shows the Li abundance 
distribution in the same bin size (0.1 dex), but shifted by 0.05 dex to smooth out the binning effect. 
The triangle in the middle panel represents the cumulative distribution of the observed Li abundance 
for the cooler stars. Results from three different sets of simulations based on the evolutionary models 
of \citet{BHAC15} are plotted by dotted ($\Delta \tau = 1$ Myr), solid ($\Delta \tau = 4$ Myr), and 
dashed ($\Delta \tau = 6$ Myr) lines, respectively. The probabilities in the right-hand panel are the 
results from the Kolmogorov-Smirnov test between the observed distribution and the synthetic distributions 
of a model cluster. Open and filled circles show the results from the evolutionary 
models of \citet{SDF00} and \citet{BHAC15}, respectively.}\label{fig8}
\end{figure}

\begin{figure}
\epsscale{0.7}
\plotone{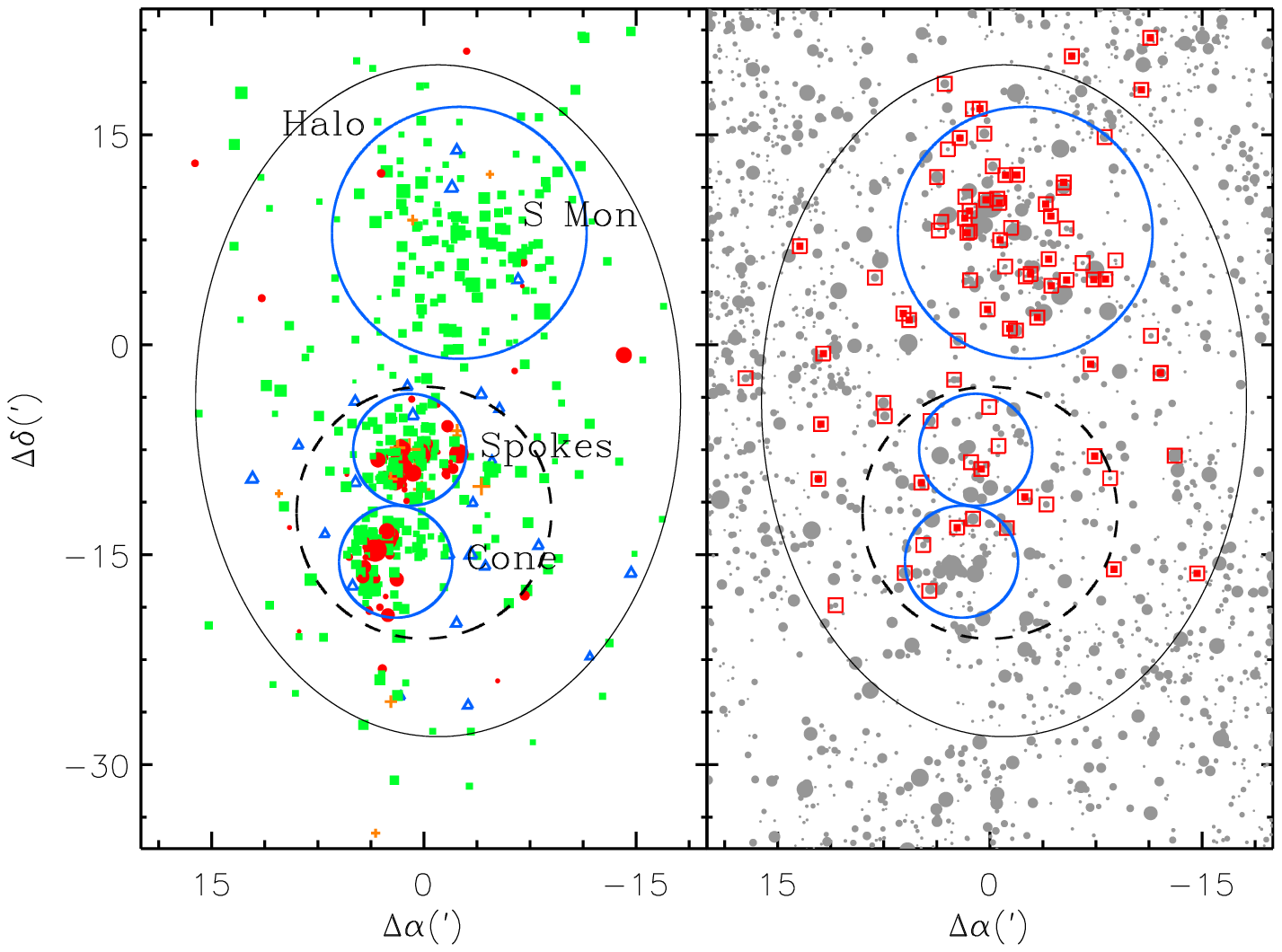}
\caption{Spatial distribution of young stellar objects (left) and stars observed in optical 
passbands (right). In the left-hand panel, filled circles (red) and filled squares (green) 
represent Class I and Class II objects, respectively. Objects with a transition disk and 
a pre-transition disk are also plotted as open triangles (blue) and crosses (orange). 
The classification of the young stellar objects was from the photometry of \citet{SSB09} for 
{\it Spitzer} Infrared Array Camera four channels and Multiband Imaging Photometer 24 $\micron$ band.  
The right-hand panel shows stars brighter than $V = 17$ mag in the photometry of 
\citet{SBCKI08}. A total 86 PMS stars are marked as open squares, and double squares represent 
41 stars in the subsample. Large open circles in both panels indicate the areas of the subclusters 
in S Mon, Spokes, and Cone nebula. An ellipse and dashed line circle are the halo and mini 
halo surrounding Spokes and Cone nebula 
regions. }\label{fig9}
\end{figure}

\begin{figure}
\epsscale{0.50}
\plotone{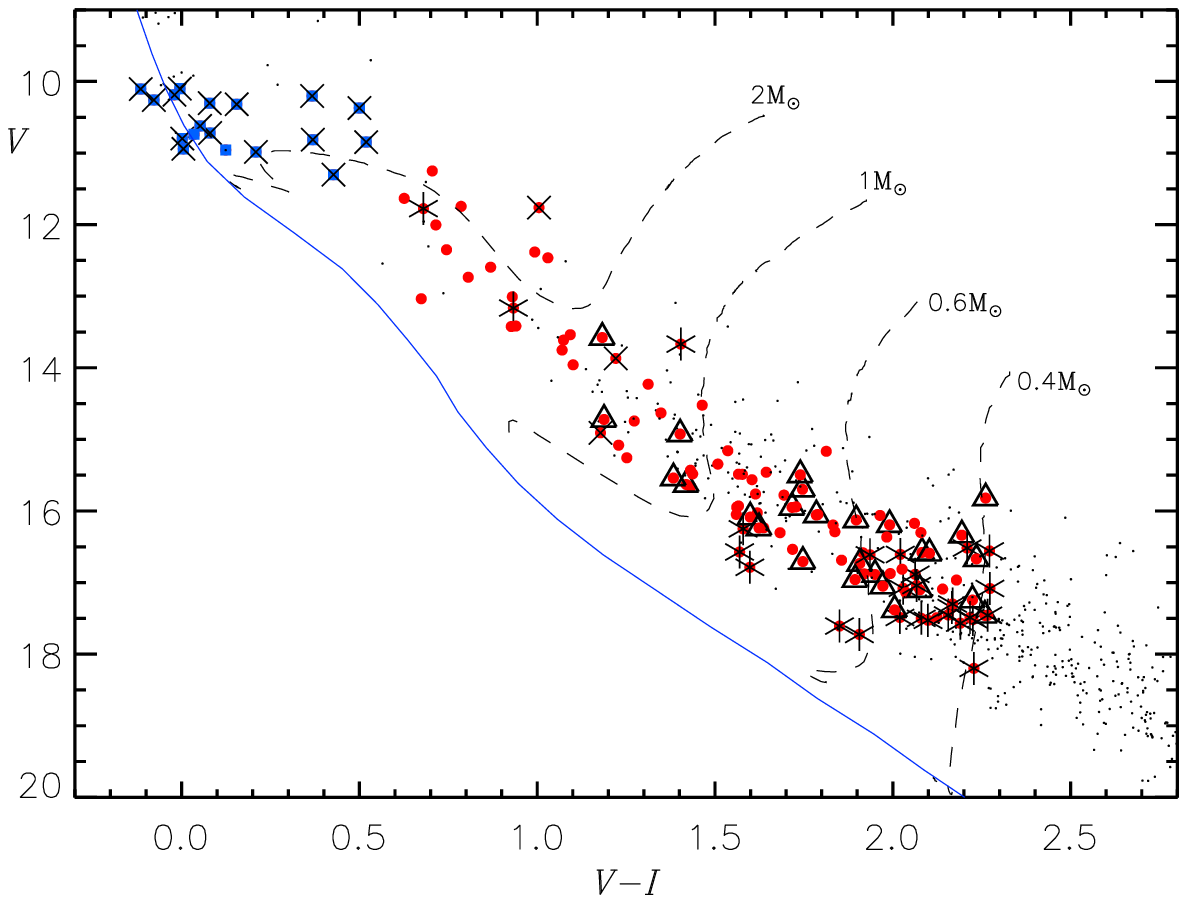}
\caption{The ($V, V-I$) color-magnitude diagram of known pre-main sequence members of NGC 2264. 
Bold dots, squares, crosses, and triangle represent the observed stars in this study, stars hotter than 6500 $K$, 
stars without Li {\scriptsize \textsc{I}} $\lambda$6708 doublet, and accreting stars, respectively. 
Stars either with poor signal-to-noise ratio or with high uncertainty in equivalent width of Li are superposed by 
an asterisk. Other symbols are the same as the right-hand panel of Figure~\ref{fig1}.}\label{fig10}
\end{figure}

\begin{figure}
\epsscale{1.00}
\plotone{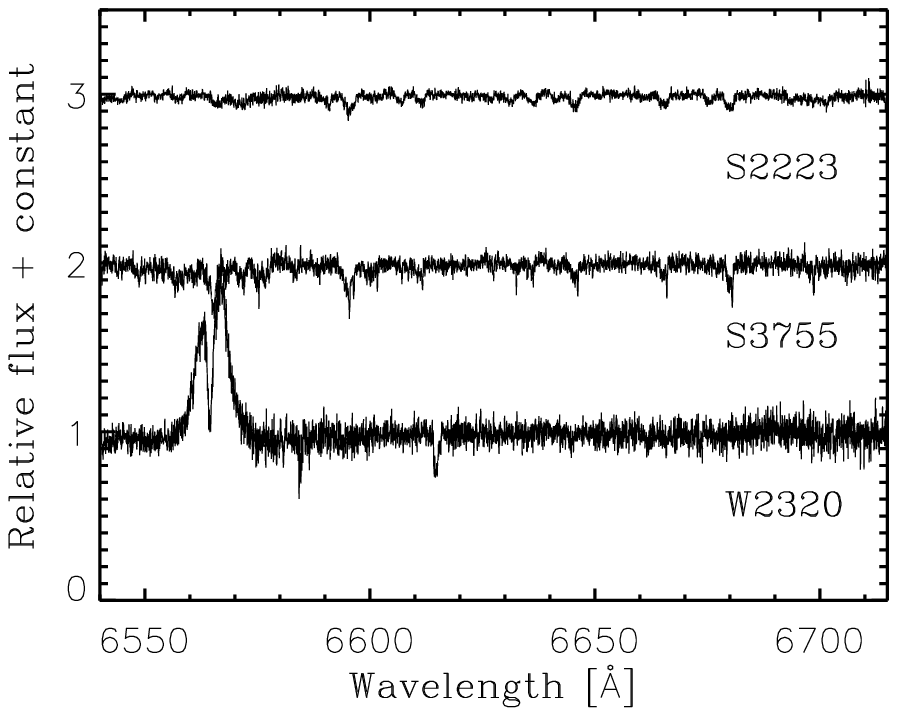}
\caption{Spectra of three Li deficient stars. The target name is labeled below each spectrum. }\label{fig11}
\end{figure}

\begin{figure}
\epsscale{0.5}
\plotone{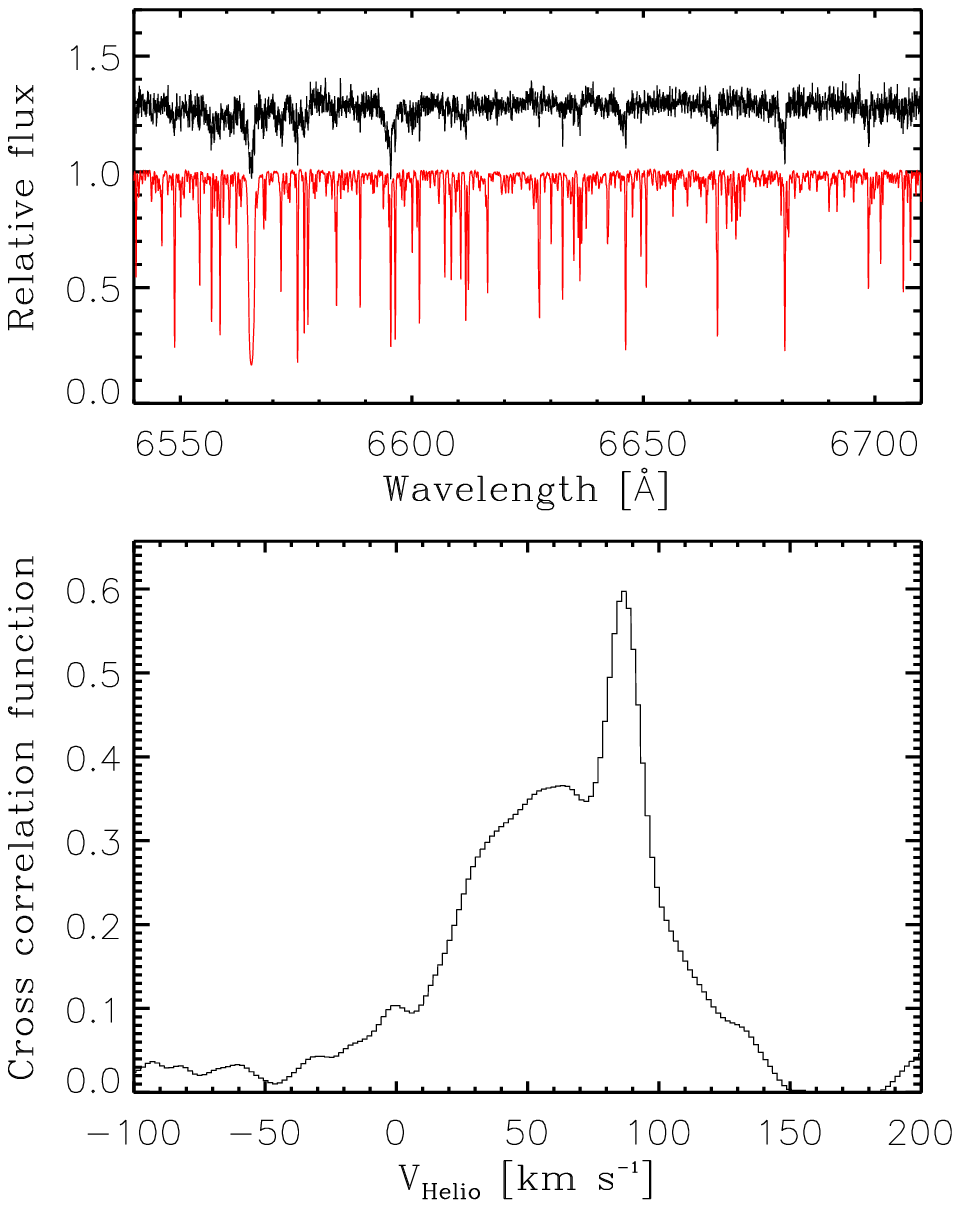}
\caption{Cross correlation between the spectra of S3755 and HD 320868 (K5). The spectra of the 
stars are plotted from top (S3755) to bottom (HD 320868) in the upper panel. The lower panel 
shows the cross correlation function between them. }\label{fig12}
\end{figure}

\clearpage









\clearpage

\begin{table}
\begin{center}
\caption{A list of field stars and additional correction values \label{tab1}}
\begin{tabular}{llcc}
\tableline\tableline
Object & Field star name & $W_{6705.1+6710.3}$(Fe {\scriptsize \textsc{I}}) (m\AA) & Adopted value (m\AA) \\
\tableline
S1776  & HD 14802 (G1V) &  127 & $125 \pm 30$ \\
            &HD 20807 (G0V) &  102 &   \\
            & HD 211415 (G0V) &  105 & \\
            & HD 216435 (G0V) &  168 & \\
S2663 & HD 100623 (K0V) &  144& 144 \\ 
S2715 & HD 10361 (K2V) &  200& $212\pm12$ \\
           & HD 22049 (K2V) &  225 & \\
\tableline
\end{tabular}
\end{center}
\end{table}

\begin{deluxetable}{lcccccccccc}
\tabletypesize{\tiny}
\tablewidth{0pt}
\tablecaption{Photometric and spectroscopic data of 86 pre-main sequence stars \label{tab2}}
\tablehead{\colhead{ID} & \colhead{$\alpha_{2000}$} & \colhead{$\delta_{2000}$} & \colhead{$V$\tablenotemark{1}} & \colhead{$V-I$\tablenotemark{1}} &
\colhead{$W_{\lambda6708}$(Li)\tablenotemark{2}}  & \colhead{$W_{\lambda6708}$(Li)$_{\mathrm{cor}}$\tablenotemark{3}} &
\colhead{A(Li)} & \colhead{$W_{\lambda6563}$(H$\alpha$)\tablenotemark{4}} &
\colhead{10\% width} & \colhead{$W_{\lambda6678}$(He {\tiny \textsc{I}})} \\
\colhead{} & \colhead{(h:m:s)} & \colhead{($\degr:\arcmin:\arcsec$)} &  \colhead{(mag)} & \colhead{(mag)} & \colhead{(\AA)} &\colhead{(\AA)} &
\colhead{} & \colhead{(\AA)} & \colhead{(km s$^{-1}$)} & \colhead{(\AA)} }
\startdata
W1322  &  06:39:34.42  & +09:54:51.2  & 16.125  & 1.897  & 0.331  &  0.305  &  1.68  &  -7.651 &  356.4  & -0.113 \\
 W2690  &  06:39:59.25  & +09:27:24.6  & 16.238  & 1.624  & 0.464  &  0.441  &  2.79  &  -4.453 &  530.6  &  0.122 \\
 S550   &  06:40:05.65  & +09:35:49.1  & 11.250  & 0.705  & 0.160  &  0.152  &  3.17  & \nodata & \nodata &  0.063 \\
 S643   &  06:40:09.59  & +09:41:43.5  & 13.417  & 0.940  & 0.246  &  0.232  &  2.85  & \nodata & \nodata &  0.058 \\
 C17089 &  06:40:09.65  & +10:08:59.3  & 16.965  & 2.179  & 0.677  &  0.650  &  3.56  &   -0.892 &  145.7  &  0.115 \\
 S645   &  06:40:09.71  & +09:41:43.6  & 16.048  & 1.560  & 0.389  &  0.366  &  2.37  &   2.537 &  170.3  &  0.128 \\
 S700   &  06:40:12.43  & +09:44:23.3  & 16.302  & 2.079  & 0.595  &  0.568  &  3.28  &   3.497 &  181.2  &  0.112 \\
 W3407  &  06:40:12.58  & +10:05:40.6  & 15.496  & 1.740  & 0.580  &  0.556  &  3.29  &   0.579 &  217.2  & -0.014 \\
 C19191 &  06:40:15.16  & +10:01:57.9  & 16.959  & 1.894  & 0.529  &  0.503  &  3.04  &   1.831 &  387.9  & \nodata\\
 C21959 &  06:40:22.62  & +09:49:46.3  & 17.142  & 2.037  & 0.491  &  0.465  &  2.77  &   3.124 &  142.0  &  0.104 \\
 S1039  &  06:40:23.08  & +09:27:42.3  & 17.382  & 2.005  & 0.439  &  0.413  &  2.44  & -14.627 &  564.7  & -0.225 \\
 C22501 &  06:40:24.16  & +09:34:12.5  & 16.707  & 1.747  & 0.313  &  0.289  &  1.68  &  -8.072 &  357.8  & -0.348 \\
 S1163  &  06:40:25.47  & +09:48:26.0  & 15.663  & 1.432  & 0.505  &  0.484  &  3.07  &   2.321 &  202.4  &  0.163 \\
 S1177  &  06:40:25.70  & +09:58:35.6  & 13.537  & 1.093  & 0.390  &  0.374  &  3.18  & \nodata & \nodata &  0.195 \\
 S1329  &  06:40:28.59  & +09:35:47.5  & 16.062  & 1.964  & 0.592  &  0.566  &  3.29  &   1.193 &  178.4  &  0.131 \\
 S1344  &  06:40:28.82  & +09:48:24.4  & 16.026  & 1.619  & 0.515  &  0.492  &  3.07  &   2.554 &  184.8  &  0.228 \\
 C24554 &  06:40:29.77  & +09:42:21.4  & 17.050  & 1.972  & 0.503  &  0.477  &  2.87  &  -5.798 &  526.6  & -0.132 \\
 C25335 &  06:40:32.01  & +09:49:35.5  & 16.887  & 1.950  & 0.316  &  0.290  &  1.54  & -62.997 &  394.3  & -0.671 \\
 W4425  &  06:40:35.11  & +10:04:21.8  & 15.167  & 1.813  & 0.499  &  0.474  &  2.91  &   2.646 &  212.2  & \nodata\\
 S1729  &  06:40:36.62  & +09:48:22.8  & 15.763  & 1.614  & 0.532  &  0.509  &  3.14  &   4.264 &  180.3  &  0.193 \\
 S1731  &  06:40:36.65  & +09:52:03.2  & 16.057  & 1.785  & 0.298  &  0.273  &  1.54  & -23.148 &  454.8  & -0.424 \\
 C27301 &  06:40:37.46  & +09:55:21.2  & 16.304  & 1.683  & 0.557  &  0.533  &  3.22  &   5.061 &  124.7  &  0.178 \\
 S1776  &  06:40:37.48  & +09:54:57.9  & 11.745  & 0.786  & 0.261  &  0.126  &  2.81  & \nodata & \nodata & \nodata\\
 S1943  &  06:40:41.02  & +09:47:57.5  & 15.950  & 1.717  & 0.527  &  0.503  &  3.09  & -14.621 &  408.0  &  0.115 \\
 S1952  &  06:40:41.12  & +09:52:56.5  & 15.698  & 1.746  & 0.506  &  0.482  &  2.97  & -10.120 &  259.0  & -0.230 \\
 S1990  &  06:40:41.78  & +09:49:52.1  & 16.290  & 1.837  & 0.562  &  0.537  &  3.20  &   3.101 &  243.3  &  0.130 \\
 S2030  &  06:40:42.45  & +09:32:20.5  & 15.257  & 1.252  & 0.426  &  0.407  &  2.99  & \nodata & \nodata &  0.241 \\
 C29037 &  06:40:42.62  & +09:53:47.6  & 16.235  & 1.637  & 0.493  &  0.470  &  2.95  & \nodata & \nodata &  0.202 \\
 S2169  &  06:40:45.04  & +09:45:41.7  & 16.197  & 1.832  & 0.644  &  0.619  &  3.50  &   1.413 &  220.4  &  0.172 \\
 C30470 &  06:40:46.95  & +09:48:48.1  & 16.536  & 1.718  & 0.489  &  0.465  &  2.89  &   4.817 &  129.3  &  0.136 \\
 C30909 &  06:40:48.38  & +09:48:38.6  & 17.459  & 2.259  & 0.516  &  0.489  &  2.81  & -12.037 &  449.8  & -0.196 \\
 C30991 &  06:40:48.62  & +09:32:52.4  & 16.879  & 1.920  & 0.464  &  0.438  &  2.65  &   2.333 &  181.4  &  0.243 \\
 S2468  &  06:40:50.85  & +09:55:53.2  & 15.563  & 1.604  & 0.559  &  0.536  &  3.24  &   2.182 &  204.0  &  0.107 \\
 S2481  &  06:40:51.18  & +09:44:46.0  & 12.004  & 0.715  & 0.119  &  0.110  &  2.94  & \nodata & \nodata &  0.151 \\
 S2528  &  06:40:52.55  & +09:52:05.9  & 16.338  & 2.194  & 0.660  &  0.633  &  3.50  &   1.811 &  407.9  & \nodata\\
 S2541  &  06:40:52.92  & +09:44:54.3  & 15.459  & 1.644  & 0.571  &  0.547  &  3.28  &  -0.855 &  213.8  &  0.097 \\
 S2568  &  06:40:53.77  & +09:30:39.1  & 11.633  & 0.626  & 0.136  &  0.129  &  3.30  & \nodata & \nodata & \nodata\\
 C32567 &  06:40:54.20  & +09:55:52.0  & 16.687  & 1.856  & 0.520  &  0.495  &  3.01  &   3.783 &  236.0  &  0.142 \\
 C32585 &  06:40:54.26  & +09:49:20.4  & 16.580  & 2.082  & 0.432  &  0.405  &  2.34  & -45.754 &  405.1  & -0.285 \\
 S2630  &  06:40:55.72  & +09:51:13.7  & 15.488  & 1.577  & 0.558  &  0.535  &  3.25  &   2.837 &  238.9  &  0.143 \\
 C33005 &  06:40:55.91  & +09:53:53.3  & 16.585  & 1.916  & 0.607  &  0.581  &  3.35  &   4.735 &  162.2  &  0.123 \\
 C33066 &  06:40:56.16  & +09:36:30.8  & 17.245  & 2.224  & 0.300  &  0.273  &  1.14  &  -9.741 &  575.1  &  0.042 \\
 S2663  &  06:40:56.50  & +09:54:10.4  & 12.383  & 0.993  & 0.532  &  0.373  &  3.40  & \nodata & \nodata & \nodata\\
 S2715  &  06:40:57.83  & +09:56:30.1  & 13.752  & 1.070  & 0.585  &  0.357  &  3.14  & \nodata & \nodata & \nodata\\
 C33749 &  06:40:58.82  & +09:39:18.7  & 17.101  & 2.076  & 0.599  &  0.572  &  3.29  &  -5.458 &  295.8  &  0.063 \\
 S2767  &  06:40:59.31  & +09:46:16.4  & 16.194  & 1.991  & 0.646  &  0.620  &  3.48  &  -6.531 &  236.2  & -0.095 \\
 S2784  &  06:40:59.74  & +09:54:06.1  & 15.942  & 1.728  & 0.497  &  0.473  &  2.93  &   4.228 &  241.4  &  0.162 \\
 S2795  &  06:41:00.25  & +09:58:49.9  & 13.613  & 1.074  & 0.365  &  0.349  &  3.09  & \nodata & \nodata & \nodata\\
 C34298 &  06:41:01.12  & +09:34:52.3  & 16.735  & 1.907  & 0.553  &  0.527  &  3.15  &  -3.217 &  300.1  &  0.076 \\
 W5413  &  06:41:01.51  & +10:14:56.0  & 16.365  & 1.983  & 0.528  &  0.502  &  3.01  & -63.831 &  211.7  & \nodata\\
 S2828  &  06:41:01.54  & +10:00:36.7  & 15.345  & 1.508  & 0.545  &  0.523  &  3.22  &   4.736 &  177.3  &  0.149 \\
 S2890  &  06:41:03.49  & +09:31:18.6  & 12.735  & 0.806  & 0.207  &  0.196  &  3.08  & \nodata & \nodata & \nodata\\
 C34897 &  06:41:03.57  & +10:00:35.4  & 16.814  & 2.026  & 0.431  &  0.405  &  2.37  &   3.746 &  139.4  &  0.099 \\
 S2912  &  06:41:04.07  & +09:35:21.2  & 16.670  & 2.235  & 0.465  &  0.438  &  2.48  &   1.195 &  289.5  &  0.056 \\
 S2921  &  06:41:04.31  & +09:48:21.6  & 14.743  & 1.273  & 0.521  &  0.502  &  3.29  & \nodata & \nodata &  0.180 \\
 S2924  &  06:41:04.41  & +09:51:50.0  & 12.464  & 1.030  & 0.384  &  0.369  &  3.27  & \nodata & \nodata &  0.137 \\
 S2928  &  06:41:04.47  & +09:53:18.4  & 16.050  & 1.789  & 0.606  &  0.581  &  3.37  &   4.204 &  187.0  &  0.110 \\
 S2949  &  06:41:05.10  & +09:51:44.5  & 15.486  & 1.566  & 0.585  &  0.562  &  3.34  &   1.102 &  151.4  &  0.203 \\
 S2967  &  06:41:05.69  & +09:54:18.8  & 16.173  & 2.061  & 0.722  &  0.695  &  3.72  &   1.666 &  222.6  & \nodata\\
 S2974  &  06:41:05.82  & +09:52:47.8  & 15.779  & 1.694  & 0.658  &  0.634  &  3.56  &   5.400 &  119.4  &  0.143 \\
 S3018  &  06:41:07.27  & +09:58:31.4  & 15.158  & 1.536  & 0.607  &  0.585  &  3.42  &   2.747 &  211.6  &  0.123 \\
 S3035  &  06:41:07.77  & +09:44:02.8  & 13.424  & 0.927  & 0.336  &  0.323  &  3.33  & \nodata & \nodata &  0.157 \\
 S3039  &  06:41:08.03  & +09:30:40.7  & 15.630  & 1.419  & 0.476  &  0.455  &  2.93  &   0.685 &  348.1  &  0.218 \\
 S3061  &  06:41:08.91  & +09:41:14.8  & 15.081  & 1.229  & 0.477  &  0.459  &  3.25  &   2.819 &  200.0  &  0.176 \\
 C36368 &  06:41:10.71  & +09:57:42.6  & 17.091  & 2.140  & 0.599  &  0.572  &  3.28  &   0.404 &  130.6  &  0.111 \\
 S3126  &  06:41:11.59  & +10:02:23.6  & 14.229  & 1.312  & 0.498  &  0.478  &  3.10  &   4.022 &  202.7  &  0.101 \\
 S3149  &  06:41:12.59  & +09:52:31.3  & 13.577  & 1.183  & 0.458  &  0.440  &  3.28  &  -5.182 &  453.5  & \nodata\\
 C36877 &  06:41:13.31  & +09:51:54.4  & 17.490  & 2.124  & 0.792  &  0.765  &  3.93  &   5.021 &  129.2  &  0.158 \\
 S3184  &  06:41:13.83  & +09:55:44.1  & 12.350  & 0.745  & 0.197  &  0.188  &  3.22  & \nodata & \nodata &  0.102 \\
 S3221  &  06:41:15.70  & +09:38:18.2  & 13.035  & 0.674  & 0.102  &  0.094  &  2.97  & \nodata & \nodata &  0.109 \\
 S3226  &  06:41:16.01  & +09:26:09.5  & 16.592  & 2.103  & 0.124  &  0.097  & -0.78  & -37.727 &  355.4  & -1.595 \\
 S3267  &  06:41:17.72  & +09:29:26.6  & 15.820  & 2.261  & 0.706  &  0.679  &  3.66  &   1.154 &  305.8  & \nodata\\
 S3282  &  06:41:18.28  & +09:33:53.6  & 14.926  & 1.402  & 0.540  &  0.519  &  3.22  &  -9.609 &  283.3  &  0.046 \\
 S3367  &  06:41:21.80  & +09:45:30.9  & 15.431  & 1.431  & 0.472  &  0.451  &  2.90  &   3.676 &   90.7  &  0.209 \\
 S3404  &  06:41:23.05  & +09:27:26.6  & 12.594  & 0.869  & 0.269  &  0.257  &  3.16  & \nodata & \nodata &  0.081 \\
 C39235 &  06:41:23.46  & +09:45:58.6  & 16.874  & 1.993  & 0.533  &  0.507  &  3.03  &   5.011 &  197.0  &  0.101 \\
 S3634  &  06:41:28.77  & +09:38:38.8  & 13.958  & 1.101  & 0.378  &  0.362  &  3.10  & \nodata & \nodata &  0.161 \\
 S3650  &  06:41:29.17  & +09:39:35.9  & 13.008  & 0.930  & 0.347  &  0.333  &  3.37  & \nodata & \nodata &  0.173 \\
 S3734  &  06:41:31.62  & +09:48:32.8  & 14.631  & 1.348  & 0.550  &  0.530  &  3.27  &   2.333 &  138.8  &  0.148 \\
 W6655  &  06:41:42.88  & +09:25:08.3  & 16.086  & 1.599  & 0.293  &  0.270  &  1.65  & -14.291 &  334.9  & -0.342 \\
 S4400  &  06:41:46.44  & +09:43:07.0  & 15.947  & 1.561  & 0.579  &  0.556  &  3.32  &   3.283 &  195.4  &  0.166 \\
 S4423  &  06:41:47.11  & +09:38:04.6  & 15.930  & 1.566  & 0.560  &  0.537  &  3.25  &   2.388 &  113.6  &  0.202 \\
 S4442  &  06:41:47.80  & +09:34:09.5  & 14.522  & 1.464  & 0.541  &  0.519  &  3.20  &   3.963 &  212.8  &  0.162 \\
 W6920  &  06:41:49.17  & +10:19:35.2  & 15.540  & 1.383  & 0.537  &  0.516  &  3.21  & -15.098 &  295.7  & -0.114 \\
 W7105  &  06:41:53.16  & +09:50:47.5  & 15.482  & 1.437  & 0.528  &  0.507  &  3.16  &   3.832 &  268.7  &  0.169 \\
 S5192  &  06:42:08.72  & +09:41:21.3  & 14.722  & 1.188  & 0.466  &  0.448  &  3.30  & -33.405 &  489.4  &  0.075 \\
\enddata
\tablenotetext{1}{The photometric data from \citet{SBCKI08}.}
\tablenotetext{2}{Equivalent width of Li {\tiny \textsc{I}} $\lambda$6708 doublet blended with adjacent neutral iron lines} 
\tablenotetext{3}{Deblended equivalent width of Li {\tiny \textsc{I}} $\lambda$6708 doublet}
\tablenotetext{4}{Full width of 10\% of peak H$\alpha$ flux}
\end{deluxetable}
\clearpage

\begin{deluxetable}{cccc}
\tablewidth{0pt}
\tablecaption{Results of Monte-Carlo simulations with different observational errors \label{tab3}}
\tablehead{\colhead{1$\sigma$ error (dex)} & \colhead{$\Delta \tau _{\mathrm{Best} }$\tablenotemark{1} (Myr)} 
& \colhead{Prob\tablenotemark{2} (\%)} & \colhead{Evolutionary model\tablenotemark{3}}} 
\startdata
0.05  & 6 & 4.0 & B \\
         & 6 & 2.5 & S \\
0.10 & 5 & 42.1 & B\\
        & 4 & 27.7 & S\\
0.15 & 4 & 66.1 & B \\
        & 4 & 73.4 & S \\
0.20 & 4 & 96.5 & B \\
        & 3 & 94.9 & S \\
0.30  & 1 & 51.5 & B\\
       & 1 & 38.2 & S \\
\enddata
\tablenotetext{1}{$\Delta \tau _{\mathrm{Best}}$ is the underlying age spread of the most probable model cluster.}
\tablenotetext{2}{Probabilities in percentile are from the Kolmogorov--Smirnov test.}
\tablenotetext{3}{B and S represent the evolutionary models of \citet{BHAC15} and \citet{SDF00} for pre-main sequence stars, respectively. }
\end{deluxetable}
\clearpage

\begin{table}
\begin{center}\scriptsize
\caption{A list of lithium deficient stars \label{tab4}}
\begin{tabular}{lccccccc}
\tableline\tableline
Object & $\alpha_{2000}$ & $\delta_{2000}$ & $V$ & $V-I$ & Membership\tablenotemark{a} & $V_{\mathrm{Helio}}$ [km s$^{-1}$]\tablenotemark{b} & Remarks \\
\tableline
S2223  & 06:40:46.07 & +09:49:17.2 &11.76 & 1.00 & X-ray emission\tablenotemark{c} & 15.6 -- 69.5 & Double lined binary\tablenotemark{b}\\
S3755   & 06:41:32.07 & +10:01:04.9 &13.87 & 1.22 & X-ray emission\tablenotemark{d} &  &  Double lined binary \\      
W2320 &06:39:51.90 & +10:12:39.4 & 14.91 & 1.18 & H$\alpha$ emission candidate\tablenotemark{e} && Double peaked H$\alpha$ emission line \\
             & & &          &         &  && Interstellar absorption at 6614 \AA \\
\tableline
\end{tabular}
\tablenotetext{a}{Membership selection criteria.}
\tablenotetext{b}{Heliocentric radial velocity and duplicity information from \citet{FHS06}.}
\tablenotetext{c}{{\it Chandra} X-ray source catalogue from \citet{SBC04}.}
\tablenotetext{d}{{\it XMM Newton} X-ray source catalogue from \citet{DSPP07}}
\tablenotetext{e}{Probable H$\alpha$ emission star from \citet{SBCKI08}.}
\end{center}
\end{table}



\end{document}